\documentclass[amsmath,amssymb,showpacs,preprintnumbers,twocolumn]{revtex4}
\usepackage[english,dutch]{babel}
\usepackage{graphicx}
\usepackage{dcolumn}
\usepackage{bm}
\usepackage{subfigure}
\newlength{\dfltgraph}

\renewcommand{\vec}[1]{\mathbf{#1}}

\newcommand \be{\begin{equation}}
\newcommand \ee{\end{equation}}
\newcommand \ba{\begin{eqnarray}}
\newcommand \ea{\end{eqnarray}}
\def\nn{\nonumber}

\begin{document}
\selectlanguage{english}

\title[Local field approximation in streamer heads]{Deviations from the local field approximation in negative streamer heads}
\author{Chao Li$^1$}
\author{W.J.M.~Brok$^2$}
\author{Ute Ebert$^{1,2}$}
\author{J.J.A.M.~van der Mullen$^2$}
\affiliation{$^1$ Centre for Mathematics and Informatics (CWI), P.O.~Box~94079, 1090~GB Amsterdam, The
Netherlands,} \affiliation{$^2$ Department of Applied Physics, Eindhoven University of Technology, P.O.~Box~513,
5600~MB, Eindhoven, The Netherlands.}


\begin{abstract}
Negative streamer ionization fronts in nitrogen under normal conditions are investigated both in a particle model and in a fluid model in local
field approximation. The parameter functions for the fluid model are derived from swarm experiments in the particle model.
The front structure on the inner scale is investigated in a 1D setting, allowing reasonable run-time and memory consumption
and high numerical accuracy without introducing super-particles. If the reduced electric field immediately before the front is
$\le 50$~kV/(cm~bar), solutions of fluid and particle model agree very well. If the field increases up to 200~kV/(cm~bar),
the solutions of particle and fluid model deviate, in particular, the ionization level behind the front becomes up to
60~\% higher in the particle model while the velocity is rather insensitive. Particle and fluid model deviate because
electrons with high energies do not yet fully run away from the front, but are somewhat ahead. This leads to increasing
ionization rates in the particle model at the very tip of the front. The energy overshoot of electrons in the leading edge
of the front actually agrees quantitatively with the energy overshoot in the leading edge of an electron swarm or avalanche
in the same electric field.
\end{abstract}

\pacs{52.80.-s,52.80.Mg,52.65.Kj,52.65.Pp}

\maketitle

\section{Introduction}\label{sec:intro}

Streamers \cite{Loe1941,Rae1964} are growing filaments of weakly ionized non-stationary plasma produced by a sharp
ionization front that propagates into non-ionized matter. Streamers are used in industrial applications such as
lighting~\cite{Lay2003,Bho2004} or gas and water purification~\cite{Vel2000,Gab2005}, and they occur in natural
processes as well such as lightning~\cite{Baz2000,Dwy2003,Wil2006} or transient luminous events in the upper
atmosphere~\cite{Sen1995}. Therefore accurate modeling and simulation of streamers is of high interest.

Most streamer models (see
e.g.~\cite{Dha1987,Guo1993,Vit1994,Kul1994,Kul1995,Bab1996,Ebe1996,Ebe1997,Mon2006:3,Seg2006,Luq2007}) are so
called fluid models for the densities of different particle species in the discharge. These models build on the
assumption of local equilibrium: transport and reaction coefficients in the continuity equations are functions of
local parameters only. If this parameter is the local electric field, we refer to this assumption as the
\emph{local field approximation}. This assumption is commonly considered to be valid as long as equilibration
length or time scales are much smaller than the spatial or temporal gradients in the electric field. For the
strongly varying electric fields within a streamer ionization front, the validity of the local field approximation
was investigated in~\cite{Nai1997,Kun1988:1,Liu2004}. The general `sentiment' in these studies is that the
approximation suffices for practical purposes and that more detailed methods tracking the behavior of individual
particles lead to just minor corrections.

Another recent result supporting the fluid approximation for streamers was that even streamer
branching~\cite{Vel2002,Brie2005,Ebe2006} can be understood in terms of an inherent instability of the fully
deterministic fluid equations~\cite{Arr2002,Roc2002,Ebe2006,Mon2006:3,Mon2006:2}. These studies have shown that
a streamer in nitrogen can reach a state in which the width of the space charge layer that creates the field
 enhancement at the streamer tip, is much smaller than the streamer diameters; the streamer then can branch
 spontaneously due to a Laplacian interfacial instability.

However, despite success and progress of fluid approximations and simulations for streamers, there are three major
reasons to reinvestigate the local field approximation:
\begin{itemize}
\item[1.] Not all streamers are alike. Experiments as well as simulations show that rapidly applied high electric
voltages can create streamers that are more than an order of magnitude faster and wider than streamers at lower
voltages~\cite{BRI2006}. Whether earlier findings on streamers in lower potentials apply to those fast and wide
streamers as well has to be investigated.
\item[2.] The detection of x-rays emanating from lightning strokes~\cite{Dwy2004,Smi2005,Dwy2005,Mos2006} indicates
that electrons can gain very high energies within early stages of the lightning event. Therefore runaway electrons
within streamer and leader processes should be investigated. Runaway electrons by definition violate a local approximation.
\item[3.] Streamer branching is an inherent instability of a fully deterministic fluid model. However, fluctuations
of particle densities might trigger this instability earlier than they would occur in the fully deterministic fluid model.
In particular, in the leading edge of an ionization front, particle densities are very low and the fluid approximation
eventually breaks down. As the front velocity of this so-called pulled front~\cite{Ebe1996,Ebe1997,Ebe2000} is determined
precisely in the leading edge region, single particle dynamics and fluctuations should be accounted for.
\end{itemize}

These three observations motivate our present reinvestigation of the local field approximation for streamers. The
starting point is a Monte Carlo model for the motion of single free electrons in nitrogen. We note that complete streamers
have been simulated with Monte Carlo particle models before~\cite{Oli2005}, however, a drawback of such models is
that the computation time grows with the number of particles and eventually exceeds the CPU space of any computer.
This difficulty is counteracted by using superparticles carrying charge and mass of many physical particles; but
superparticles in turn create unphysical fluctuations and stochastic heating.

\begin{figure}
\begin{center}
\includegraphics[width=0.5\textwidth]{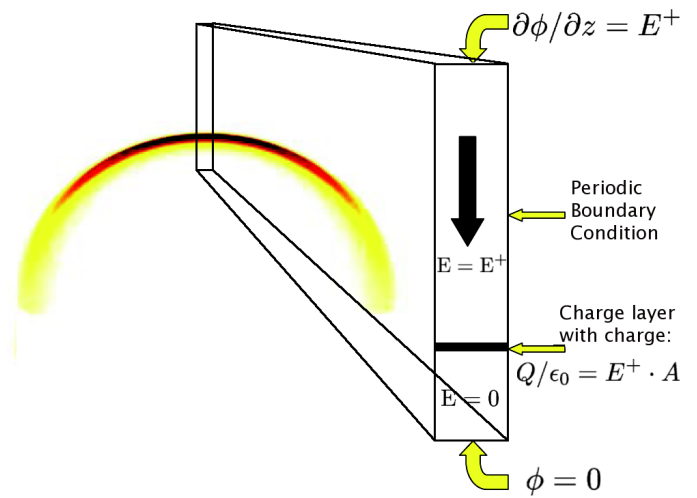}
\end{center}
\caption{\label{fig:planarfront}(Color online) 
The relation between the full streamer problem and the planar
fronts described in this paper: the left picture shows the narrow space charge layer surrounding the streamer
head~\cite{Roc2002, Mon2006:3, Mon2006:2}; the width of the layer is much smaller than the streamer diameter which creates the characteristic field
enhancement ahead  and field suppression behind the front. The right picture shows a zoom into the inner structure
of the space charge layer with an essentially planar ionization front as treated in this paper. In the transversal
direction, periodic boundary conditions are applied.}
\end{figure}

In the present paper, we compare the results of a Monte Carlo particle model and a fluid model. We circumvent the
problems caused either by a too large particle number or by the introduction of superparticles by investigating a
small, essentially one-dimensional section of the ionization front as illustrated in Fig.~\ref{fig:planarfront}.
We suppress effects of lateral boundaries by periodic boundary conditions. As the electric field essentially does
not deviate from the planar geometry within the region where the particle densities vary rapidly, a planar
ionization front~\cite{Lag1994,Ebe1996} is a very good approximation of this inner structure. Of course, a planar
front will not incorporate the electric field enhancement caused by a curved front~\cite{Dha1987,Lag1994}, but
this outer scale problem concerns only the electric field and can be dealt with through an inner-outer
matching procedure~\cite{Ben1978,Fif1988,Ebe2000:1}. Planar fronts allow us to investigate individual
particle kinetics and fluctuations within the front and its specific strong spatio-temporal gradients in a
systematic way and within reasonable computing time.

In this paper, we concentrate on negative streamer fronts in pure nitrogen under normal conditions. We thoroughly discuss the case
where the reduced electric field at the streamer tip is 100~kV/(cm~bar), and we summarize
results for fields ranging from 50 to 200~kV/(cm~bar). The paper is organized as follows. In
section~\ref{sec:model}, first our Monte Carlo particle code and its numerical implementation are described.
Then the derivation of the fluid model is recalled, and the numerical implementation of the fluid model
is summarized. Then swarm or avalanche experiments in a fixed field are performed in the particle model,
the approach of electrons to a steady state velocity distribution is investigated, and the parameter functions
for the fluid model are generated. This sets the stage for a quantitative comparison of front solutions in
particle and fluid model in section~\ref{sec:result}. Here first the setup of planar
front simulations is described, then the results of the planar front simulations within fluid and particle model
and analytical results are presented and compared. The emphasis lies on front profile, front velocity and ionization
level behind the front. It will be shown that differences can be attributed to the electron kinetics in the leading
edge of the front where the electric field does not vary, and that the electron energy distribution there agrees
quantitatively with that in the leading edge of an ionization avalanche or swarm.
Section~\ref{sec:conclusion} contains our conclusions on the validity of the fluid approximation.
An appendix contains analytical approximations on the ionization level behind an ionization front.

\section{Set-up of particle model and fluid model in local field approximation}\label{sec:model}

In this section, we summarize features of particle and fluid models, their numerical implemention and mutual
 relation as a basis for the quantitative comparison of ionization fronts in particle and fluid model in section~\ref{sec:result}.

Our starting point is a model that contains all microscopic physical mechanisms that are thought to be relevant
for the propagation of a negative impact ionization front in pure nitrogen. It models the generation and motion of
single free electrons and positive ions in the neutral background gas. While propagating freely, the electrons
follow a deterministic trajectory according to Newton's law. The collision of electrons with neutral molecules is
treated as a stochastic Monte Carlo process. Because the mobility of the positive ions is two orders of magnitude
smaller than that of the electrons, ions are treated as immobile within the short time scales investigated in this
paper. Neutral molecules are assumed to have a uniform density with a Maxwellian velocity distribution. The
electron-neutral collisions, including all relevant elastic, excitation, and ionization collisions, are treated
with the Monte Carlo method. Electron-electron or electron-ion processes as well as density changes of the neutral
gas are neglected as the degree of ionization stays below $10^{-5}$ even at atmospheric
pressure~\cite{Ebe1997,Ebe2006}. This well-known model will be summarized in section~\ref{sec:model_mc}. The space
charges can change the local electric field, this is accounted for by solving the Poisson equation. The particle
model gives a very detailed and complete description at the expense of significant computational costs where we
stress that one particle is one electron and superparticles are not used.

If densities are high enough and fields vary slowly in space and time, the average behavior of the electrons can
be modeled by a fluid approximation for electron and ion densities whose parameters depend on the local electric
field only. The derivation of the fluid approximation can be formalized by taking the first and the second moment
of the Boltzmann equation as is recalled in section~\ref{sec:Boltz}.
However, for the practical purpose of determining mobilities, ionization rates and diffusion
coefficients as a function of the electric field, we directly perform swarm experiments with the particle model
in a constant electric field. This procedure together with the averaging processes involved are described in
section~\ref{swarm}. Here also the relaxation of an electron swarm to steady state motion and the velocity
distribution of steady state motion in a given homogeneous field are discussed.

\subsection{The Monte Carlo particle model}\label{sec:model_mc}

\subsubsection{Physical processes}\label{sec:mc_physics}

In the particle scheme, at each instant of time $t$, there is a total number of $N_e(t)$ electrons and $N_p(t)$
ions. The single electrons are numbered by $i=1,\ldots,N_e(t)$ at time $t$; they are characterized by a position
${\bf x}_i(t)$ and a velocity ${\bf v}_i(t)$, each within a continuous three dimensional vector space. Between
collisions, electrons are accelerated and advanced according to the equation of motion
\begin{equation}
\label{eq:motion} {\rm m} \frac {\partial {\bf v}_i} {\partial t}={\rm e}{\bf E}({\bf x}_i,t), ~~~\frac{ \partial{\bf x}_i}{\partial t}={\bf v}_i,
\end{equation}
where m and e are electron mass and charge and ${\bf E}({\bf x},t)$ is the local electric field. For the positive
ions, only their position ${\bf x}_j^p$, $j=1,\ldots,N_p(t)$ is taken into account while their mobility is so low
that their velocity can be neglected. The electric field is determined from the Poisson equation
\begin{equation}
\nabla \cdot{\bf E}= \frac{{\rm e}\;(n_p-n_e)}{\epsilon_0} \label{eq:Poisson}
\end{equation}
together with appropriate boundary conditions on the electrodes. Here $n_e$ and $n_p$ are electron and ion
density, respectively. Within a particle model, these densities are
\begin{equation}
\label{density} n_e({\bf x},t)=\sum_{i=1}^{N_e(t)}\delta^3\Big({\bf x}-{\bf x}_i(t)\Big), ~n_p({\bf x},t) =
\sum_{j=1}^{N_p(t)}\delta^3\Big({\bf x}-{\bf x}_j^p(t)\Big),
\end{equation}
where $\delta$ is Dirac's $\delta$ function.

\begin{figure}
\begin{center}
\includegraphics[width=0.5\textwidth]{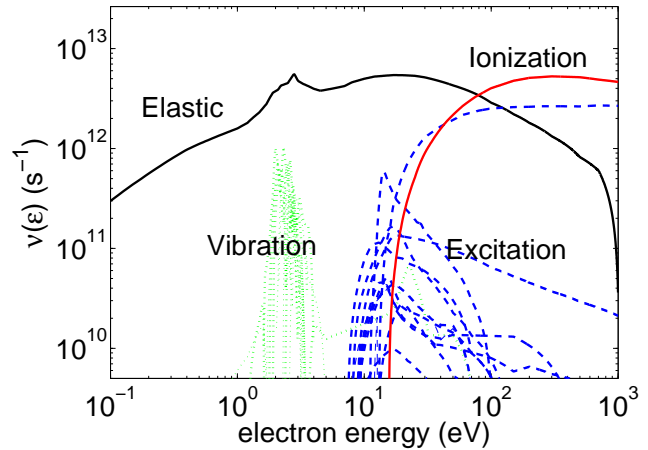}
\end{center}
\caption{\label{fig:fre_sep} Collision frequencies in N$_2$ at 1 bar from \cite{Mor1999}.}
\end{figure}

The collisions account for the impact of free electrons on neutral nitrogen molecules. As the neutrals are
abundant, their density determines the probability of an electron-neutral collision. The collision can be elastic,
inelastic or ionizing. In inelastic collisions, electron energy is partially converted into molecular excitation;
in ionizing collisions, electron energy is consumed to split the neutral into a positive ion and a second free
electron. The probability distribution of the different collision processes depends on the electron energy at the
moment of impact; we use the cross section data from the Siglo Database~\cite{Mor1999}. Fig.~\ref{fig:fre_sep}
shows the frequencies of different electron-neutral collision processes for electrons as a function of energies in
pure N$_2$ at $1$ bar pressure. As the collisions are random within a probability distribution, the actual
occurrence of a specific collision within a sample is determined by a Monte Carlo process.

Once an elastic or inelastic collision process is chosen, the energy loss of the electron and therefore the
absolute value of its velocity after the collision is fixed. However, model results will depend on the angular
distribution of the emitted electrons, which again follows a probability distribution. Different scattering
methods have been discussed in the literature \cite{Kun1986:1,Okh2002,Phe1985,Boe1982,Sur1990}. Here we will only
focus on the scattering method used in the present paper.

In an elastic collision, the longitudinal scattering angle $\chi$ and the azimuthal scattering angle $\varphi$
relative to the direction of the incident electron, are given as \cite{Okh2002}
\begin{eqnarray}\label{equ:scattering}
\cos\chi  & =  & 1- \frac{2p_1(1-\xi(\varepsilon))}{1+\xi(\varepsilon)(1-2p_1)},\\
\varphi & = & 2 \pi p_2,
\end{eqnarray}
where the random numbers $p_1$ and $p_2$ are drawn from a uniform distribution between 0 and 1, and
$\xi(\varepsilon)$ is a function of the incident particle energy $\epsilon$. The elastic cross section
data \cite{Phe1985} for nitrogen was fitted empirically in \cite{Okh2002} with
\begin{equation}\label{equ:xi}
\xi(\varepsilon)=\frac{0.065 \;\varepsilon/{\rm eV}+0.26\sqrt{\varepsilon/{\rm eV}}}
{1+0.05\;\varepsilon/{\rm eV}+0.2\sqrt{\varepsilon/{\rm eV}}}-\frac{12\sqrt{\varepsilon/{\rm eV}}}
{1+40\sqrt{\varepsilon/{\rm eV}}}.
\end{equation}

In an inelastic collision, the energy loss of incident electrons has to be taken into account,
but the scattering angle is calculated in the same way as for an elastic collision.

In an ionizing collision, energy conservation dictates
\begin{equation}\label{equ:ionenergy}
\varepsilon_1+\varepsilon_2=\varepsilon_c-\varepsilon_{ion}
\end{equation}
where $\varepsilon_c$, $\varepsilon_1$  and $\varepsilon_2$ are the energy of the incident, the scattered and the
ejected electron, respectively, and $\varepsilon_{ion}$ is the ionization threshold energy. We use Opal's empirical
fit for the distribution of the ejected electron energy \cite{opa1971}
\begin{equation}\label{equ:ionpartition}
\varepsilon_2  =  B \tan \left[p_3  \arctan \frac{\varepsilon_c-\varepsilon_{ion}}{2 B} \right],
\end{equation}
where $B \approx 13$ eV in the energy range of interest and $p_3$ again is a random number equally distributed
between 0 and 1. For the scattering angles, Boeuf and Marode \cite{Boe1982} assumed that $(i)$ the incident,
ejected and scattered electron velocities are coplanar, and $(ii)$ that the scattered and ejected electron
velocities are perpendicular. These assumptions lead to
\begin{equation}\label{equ:ionscattering}
\cos^2\chi_1=\frac{\varepsilon_1}{\varepsilon_c-\varepsilon_{ion}}, \hspace{0.5cm}
\cos^2\chi_2=\frac{\varepsilon_2}{\varepsilon_c-\varepsilon_{ion}}.
\end{equation}
where $\chi_{1,2}$ are the respective scattering angles.
The set of equations~\eqref{equ:ionenergy},~\eqref{equ:ionpartition}, and \eqref{equ:ionscattering} determines the
distribution of energies and scattering angles of the scattered and the ejected electron in an ionizing collision.

\subsubsection{Numerical implementation}
\label{sec:num-part}

The particle code moves electrons within the applied plus the self-induced field and includes their collisions.
Therefore the numerical calculation consists of three parts: the Newtonian electron motion within the field,
the field generated by the charged particles, and collisions. At each time step of length $\Delta t$, the field is calculated
from the charge densities on a lattice with grid size $\Delta \ell$. Then the electrons move in continuous phase space according
to the field, possibly interrupted by Monte Carlo collision processes. Electrons can experience more than one collision
during one time step $\Delta t$.

In more detail, position and velocity of the electrons are determined in continuous phase space from their Newtonian equation
of motion~\eqref{eq:motion} according to the electric field at their initial position within the time interval. The commonly used integration is the leap-frog method~\cite{Bir1991}, in which the electron positions and velocities are offset in
time by $\Delta t/2$.

For the electron-neutral collisions, time, type and scattering angles are determined in a Monte Carlo process by sequences
of random numbers. For the data shown in Fig.~\ref{fig:fre_sep}, the maximal collision frequency $\nu_{\rm max}$ is about
$9.7 \cdot10^{12}$/s , therefore the minimal average collision time $T_{\rm min}= 1/\nu_{\rm max} $ is about 0.1 ps.
By introducing so-called null-collisions, in which no collisions occur, $T_{\rm min}$ can be chosen as average
collision time independently of the electron energy $\varepsilon$. The probability $P(t)$ that an electron will travel a time $t$ without collision (including null collisions) is 
\begin{equation}
P(t)=e^{- \nu_{\rm max} t} .
\end{equation}
 Therefore the next collision time $\Delta t_{collision}$ of an electron is drawn in
a Monte Carlo process from the distribution
\begin{equation}
\Delta t_{collision}= T_{\rm min} \cdot \ln \frac{1}{p_4}
\end{equation}
where $p_4$ is again a random number.  
When a collision occurs, the energy of the incident electron is calculated, and the distribution of the collision
processes is determined according to the collision frequencies from Fig.~\ref{fig:fre_sep}; then a random number
determines the collision type (null collision, elastic, excitation or ionization collision). At the collision,
the electron velocities are changed according to the processes discussed in Sec.~\ref{sec:mc_physics}. Then the
next collision time for the particle is determined. This approach is described in more detail in~\cite{Boe1982,Wou2005}.

At each time step of length $\Delta t$, the electric field is calculated on the grid with mesh $\Delta \ell$. First, the number of
elementary charges $n_p-n_e$ within a grid cell is counted; it directly determines the charge density within the cell.
Then the change of electric field components normal to the cell faces are determined from the densities within the cells
 through the Poisson equation~\eqref{eq:Poisson}. This simple interpolation on cells of
appropriate size causes no artifacts as we are dealing with particles carrying just one elementary charge e, not with
superparticles. The condition on the cell size is $(i)$ that it is large enough that a single charge in the cell center
does not create substantial fields on the cell boundary, and $(ii)$ that it is small enough that no strong density
gradients occur between neighboring cells. Here it should be noted that density gradients due to particle number
fluctuations are strongly suppressed when we deal with real particles, not superparticles. Therefore more involved
interpolation methods like Particle in Cell (PIC)~\cite{Bir1991} are not required. 

The choice of the spatial and temporal mesh determines the computational accuracy as well as the computational costs.
We have tested different meshes in planar fronts as described in section~\ref{sec:con_pf}. The results, most
prominently the ionization density behind the front, converge for a sufficient discretization.
However, a balance has to be found between computational accuracy and computational costs.
We choose the time step as $\Delta t=0.3$ ps, which is of the same order as the minimal average ionization time $T_{\rm min}$,
and the cell size as $\Delta \ell= 2.3$ $\mu$m, which is the basic length scale according to dimensional analysis in~\cite{Ebe1997}.
On this mesh, the ionization density behind planar fronts has an error of less than $0.2 \%$.

\subsection{The fluid model}\label{sec:model_fl}

\subsubsection{Derivation from the Boltzmann equation}
\label{sec:Boltz}

Fluid models in general are derived from the Boltzmann equation~\cite{Hir1954,Cha1974,Shk1966}. For each species
$k$, this equation describes the temporal development of the $6$-dimensional phase space density
$f_k(\vec{x},\vec{v})$. For the electrons, omitting the subscript $k$, the Boltzmann equation reads
\begin{equation}
\frac{\partial f}{\partial t} +
        \vec{\nabla}_x \cdot \vec{v} f -
        \vec{\nabla}_v \cdot \frac{{\rm e}\vec{E}}{\rm m} f
    = \left(\frac{\delta f}{\delta t}\right)_\textsc{c}.
\label{eq:boltzmann}
\end{equation}
e$\vec{E}/m$ is the acceleration of the electrons under the influence of
the electric field. The term on the right hand side signifies the changes in the distribution function due to
collisions that happen instantaneously. It is an integral expression that contains the phase space density
$f(\vec{x},\vec{v})$ both of the electrons and of the collision partners.

Equation~\eqref{eq:boltzmann} can also be represented as an infinite series of velocity moments. For the
$l^\mathrm{th}$ moment, one multiplies the Boltzmann equation (\ref{eq:boltzmann}) with an $l^\mathrm{th}$-order
vector function $g(\vec{v}\vec{v}\vec{v}\cdots)$ and integrates over velocity. The zeroth moment results from
$g=1$, the first from $g={\rm m}\vec{v}$, the second from $g=\frac{1}{2}{\rm m}\vec{v}\vec{v}$, etc.
The resulting equations are continuity equations for density, momentum, kinetic energy,
etc.~\cite{Shk1966,Cha1974,Gog1992:1}.

Each momentum equation contains quantities that can be obtained from higher moments. Therefore the series of
momentum equations has to be truncated in a convenient way, and one needs to introduce an assumption in the
highest moment equation to be included. If we truncate after the first moment and simplify
further~\cite{Gog1992:1} in the momentum equation, we retain: \ba \frac{\partial n_e}{\partial t} + \vec{\nabla}
\cdot {\bf j}_e &=& {\cal S} \label{eq:continuity} \label{fluid1}
\\
{\bf j}_e &=& - \mu \vec{E} n_e - {\bf D} \vec{\nabla} n_e \label{eq:momentum} \ea
where $n_e$ is the electron density, ${\bf j}_e = \vec{u}n_e$ is the flux and $\vec{u} = \langle\vec{v}\rangle$ is
the mean velocity of electrons. ${\cal S}$ is the source of electrons due to collisions and impact ionization,
$\mu$ represents the mobility and ${\bf D}$ is the diffusion matrix.

The coefficients ${\cal S}$, $\mu$ and ${\bf D}$ appearing in equations~\eqref{eq:continuity} and
~\eqref{eq:momentum} are to be obtained from elsewhere. One common approach is to solve the Boltzmann
equation~\eqref{eq:boltzmann} for a homogeneous and constant electric field $E$ within a background gas of
constant density. In a uniform electric field, the electrons gain energy from the field and loose it in inelastic
collisions, reaching some steady state transport conditions~\cite{sat1985,Ale1996}. By separating
$f(\vec{x},\vec{v},t)$ in $F(\vec{v})n_e(\vec{x},t)$, expanding $F(\vec{v})$ in spherical harmonics, and
truncating this series, a practically solvable set of ordinary differential equations results~\cite{Hag2005}.
Solving these equation determines the electron velocity distribution function $f(E,\vec{v})$ from which $\mu(E)$
and ${\bf D}(E)$ with $E=|{\bf E}|$ can be calculated. Furthermore, the electron source term is written as \be
\label{S} {\cal S}=|n_e\;\mu(E)\;{\bf E}|\;\alpha(E), \ee when attachment and recombination can be neglected.
Using these coefficients in a given gas and density as a function of the electric field is called the local field
approximation.

Of course, this fluid model has to be extended by continuity equations for other relevant excited or ionized
species. For a nonattaching gas with neglected ion mobility, the continuity equation for the density $n_p$ of
positive ions has to be included
\begin{equation}
\label{eq:np}
\frac{\partial n_p}{\partial t} = {\cal S}.
\end{equation}
Alternatively, the fluid model \eqref{fluid1}--\eqref{eq:np} can also be motivated by physical considerations
and conservation laws~\cite{Ebe1997,Mon2006:1,Mon2006:3}.

In the present paper, the highest possible consistency between particle and fluid model is achieved by determining
the transport coefficients and ionization rate $\mu(E)$, matrix ${\bf D}({\bf E})$, and
$\alpha(E)$ for the fluid model from swarm experiments in the particle model; this will be done in section~\ref{swarm}.

Solutions of the particle model and of the fluid model in local field approximation will differ when the electric
field or the electron density vary rapidly in space or time as the electrons then will not fully ``equilibrate'' 
to the local electric field~\cite{Nai1997}. We will investigate these deviations further below.

\subsubsection{Numerical implementation}
\label{sec:num-fluid}

The fluid equations are solved on a uniform grid where the electron densities $n_e$ and ion densities $n_p$ are
calculated at the centers of the grid cells. The densities can be viewed as averages over the cell like in the
particle model. The field strength is also calculated at the cell centers. The electric field components are
taken on the cell faces, where they determine the mass fluxes.

The equations for the particle densities are discretized in space with the finite volume method, based on mass
balances for all cells. The particle densities are updated in time using the third order upwind-biased advection scheme
combined with a two-stage Runge-Kutta method. For the details and the tests of the algorithm, we refer
to~\cite{Mon2006:3}.

Analytical studies~\cite{Ebe1997,Ebe2000} and numerical investigations~\cite{Mon2006:2,Mon2006:3} show that the
ionization front is a pulled front; therefore a very fine numerical grid is required in the leading edge region
of the front, and standard refinement techniques refining in the interior front region fail. Like for fronts in
the particle model, we also have tested different numerical meshes for fronts in the fluid model, these fronts
are treated in section ~\ref{sec:con_pf}. On a too coarse grid, the front moves too fast and is too smooth due to
numerical diffusion of the electron density. To achieve the same numerical accuracy below 0.2~\% as for the
particle model, the fluid model requires an approximately four times finer mesh, namely $d x= 0.575$ $\mu$m
and $d t=0.075$ ps. This mesh will be used below.

\subsection{Kinetics and transport of electron swarms in constant fields}
\label{swarm}

Swarm experiments deal with electron swarms moving and multiplying in a constant electric field without
changing it. If the field is high enough such that the electron number grows measurably, such a
swarm is also called an avalanche. Swarms or avalanches in homogeneous fields are an important experimental
and theoretical tool to investigate the electron dynamics.

\subsubsection{Particle swarm kinetics: approach to steady state}

Particle swarm experiments can be used to determine transport coefficients, but also to study the particle
kinetics. We will first study the second issue, namely the relaxation of electrons to a steady state velocity
distribution and the velocity distribution itself. Indeed the electron swarm will rapidly ``equilibrate'' to the
 applied field. In such a balanced state, the electrons on average gain as much energy from the electric
field as they loose in inelastic and ionizing collisions; this is how they reach an energy and velocity
distribution specific for the electric field. The time that the electrons need to get in balance with the local
electric field is an important indication for the validity of the local field approximation. We therefore test
it here within a particle swarm experiment.

\begin{figure}
  \begin{center}
      \includegraphics[width=0.5\textwidth]{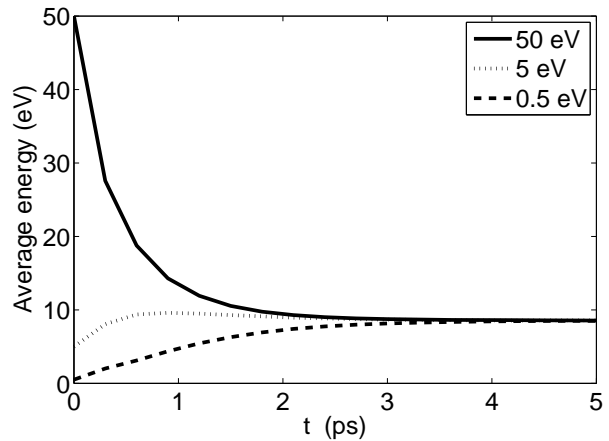}
 \end{center}
 \caption{\label{fig:en_relax} Average electron energy as a function of time for three different electron swarms
 in a field of $-100$ kV/cm. Starting with electron swarms
 moving into the drift direction with an identical kinetic energy of 50 eV (solid), 5 eV (dotted) and 0.5 eV (dashed),
 all electron swarms approach a mean electron energy characteristic for the applied field within 2 to 3 ps.}
\end{figure}

Fig.~\ref{fig:en_relax} shows how different electron swarms converge to the same mean energy within a field of $-100$
kV/cm. The experiment starts with a group of electrons of identical velocity directed in the electron drift direction;
their kinetic energy is 50, 5 and 0.5 eV, respectively. When the swarms start to drift, their average energies converge
to the same constant value within at most 3 ps in all three cases. Here the average energy is used as the
simplest indication of their energy and velocity distribution function.

\begin{figure}
 \begin{center}
           \includegraphics[width=.5\textwidth]{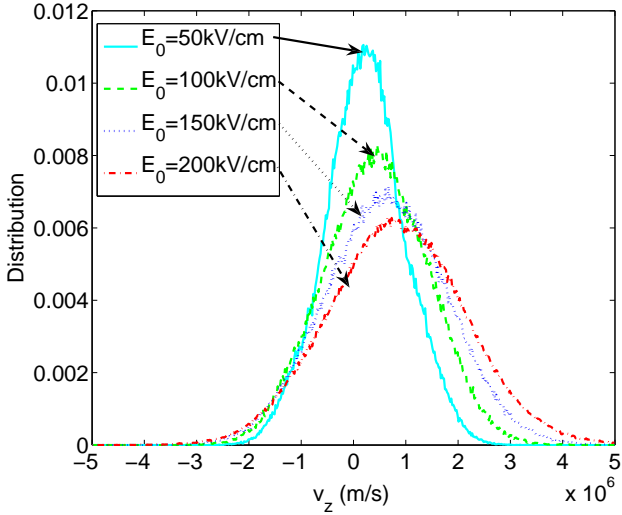}
\end{center}
     \caption{\label{fig:eedfevdf} Distribution of the electron velocity $v_z$ in the longitudinal direction in particle swarm
     experiments in fields of 50, 100, 150, and 200 kV/cm. The higher the field, the more the distribution deviates
     from Maxwellian and from symmetry about velocity $v_z=0$.}
\end{figure}

After reaching steady state motion, swarms demonstrate a steady state distribution of electron velocities and energies in
a given electric field. Fig.~\ref{fig:eedfevdf}
shows the distribution of the longitudinal electron velocity $v_z$ in an electric field of 50, 100, 150, and 200 kV/cm.
The figure shows that with increasing field, the electron velocity distribution deviates more and more from the
Maxwellian profile and therefore from symmetry about velocity $v_z=0$; rather an increasing number of electrons flies
in the direction of the field and a decreasing umber against it, and the number of electrons with high kinetic energy
increases.

\subsubsection{Gaussian swarm profiles and transport coefficients}

Swarm experiments are used as well to determine mobilities, reaction rates and diffusion constants experimentally
~\cite{Hux1974}. An electron swarm drifts, broadens and grows under the influence of a constant electric field.
The same experiment is performed here for this purpose, but now numerically with the particle model.

We here recall the essentials: A single electron will generically evolve into a swarm that has a Gaussian profile
in space. In terms of the fluid model (\ref{fluid1})--(\ref{eq:np}), this Gaussian distribution is given by
\begin{eqnarray}
\label{Gauss}
n_e(x,y,z,t)&\propto&
e^{\:\textstyle \mu E\alpha(E)\;t}\;\; \frac{e^{\textstyle -(x^2+y^2)/(4\pi D_T (t-t_0))}}{4\pi D_T (t-t_0)}
\nn\\
&& \cdot \frac{e^{\textstyle -(z-z_0-\mu Et)^2/(4\pi D_L (t-t_0))}} {\sqrt{4\pi D_L (t-t_0)}},
\end{eqnarray}
(The discharge specific context of this solution can be found in~\cite{Rai1991,Mon2006:1}.)
Here the center of the package is at $(x,y,z)=(0,0,z_0)$ at time $t=0$, and
the field is in the $z$-direction. The longitudinal and transversal components of the diffusion matrix are denoted
as $D_L$ and $D_T$.

The transport and reaction coefficients can be determined from this equation by
\begin{eqnarray}\label{equ:tof}
\mu(E)|E|&=&
\frac{\langle z(t_2)\rangle -\langle z(t_1)\rangle}{t_2-t_1},\\
\alpha(E) & = & \frac1{\mu(E)|E|}\;\frac{\ln N_e(t_2) - \ln N_e(t_1)}{t_2-t_1},\\
D_T(E) & = & \frac{\langle x^2(t_2)+y^2(t_2)\rangle -\langle x^2(t_1)+y^2(t_1)\rangle}{4(t_2-t_1)}, \\
D_L(E) & = & \frac{\langle [z(t_2)-\langle z(t_2)\rangle ]^2\rangle -\langle [z(t_1)-\langle z(t_1)\rangle
]^2\rangle }{2(t_2-t_1)},\;\; \label{tof-e}
\nonumber\\ &&
\end{eqnarray}
where the spatial averages of observables ${\cal O}$ are taken over the electron density (\ref{Gauss}) \be \langle
{\cal O}(t) \rangle = \int dx\;dy\;dz \;{\cal O}\;n_e(x,y,z,t)\quad\mbox{for the fluid model}, \ee and $N_e(t)$ is the
total number of electrons at time $t$.

When swarm experiments are performed in the particle model, transport and reaction coefficients are determined by
(\ref{equ:tof})--(\ref{tof-e}) as well, and averages are now performed over all electrons \be \langle {\cal
O}(t) \rangle= \frac1{N_e(t)}\sum_{j=1}^{N_e(t)}{\cal O}_j(t)\quad\mbox{for the particle model}, \ee in agreement with
equation (\ref{density}). This is how the coefficients for the fluid model will now be determined from the particle model.

\subsubsection{Fluid parameters determined from particle swarms}

We have determined $\mu(E)$, $D_T(E)$, $D_L(E)$, and $\alpha(E)$, and also the average electron energy $\epsilon(E)$
in particle swarm experiments for $42$ different background electric fields ranging from 2 kV/cm to 205 kV/cm.

To obtain the transport coefficients and mean values with satisfactory statistics, one needs a sufficient number of
electrons that have experienced an adequate number of collisions. The experiments start from a number of electrons at the
same position (i.e. located in a single point, which is, a Gaussian with zero width), and end with a swarm of
electrons with a Gaussian distribution as described in equation~\eqref{Gauss}. Because the ionization rate depends strongly on the electric field strength, the number of initial electrons and the simulation time is chosen according to the fields. For example, the simulation starts with $10^6$ electrons at $2$ kV/cm and lasts for $1500$ ps, but for $205$ kV/cm, the simulation starts with $10^2$ electrons and ends with $4 \cdot 10^6$ electrons after $30$ ps. 

As there is some initial transient during which the electrons ``equilibrate'' to the field and approach a Gaussian
density profile, the transport and reaction coefficients are evaluated according to
equations~\eqref{equ:tof}--\eqref{tof-e} at appropriate times $t_{1,2}$. We choose $t_2$ as the end of a swarm
experiment, and $t_1=t_2/2$ in the middle of an experiment. In view of the relaxation times below 3 ps evaluated
above, this choice of $t_1$ is on the very safe side. 

\begin{figure*}
    \centering
    \subfigure[ ~Mobility $\mu(E)$]{
          \label{fig:coe_mu}
          \includegraphics[width=.47\textwidth]{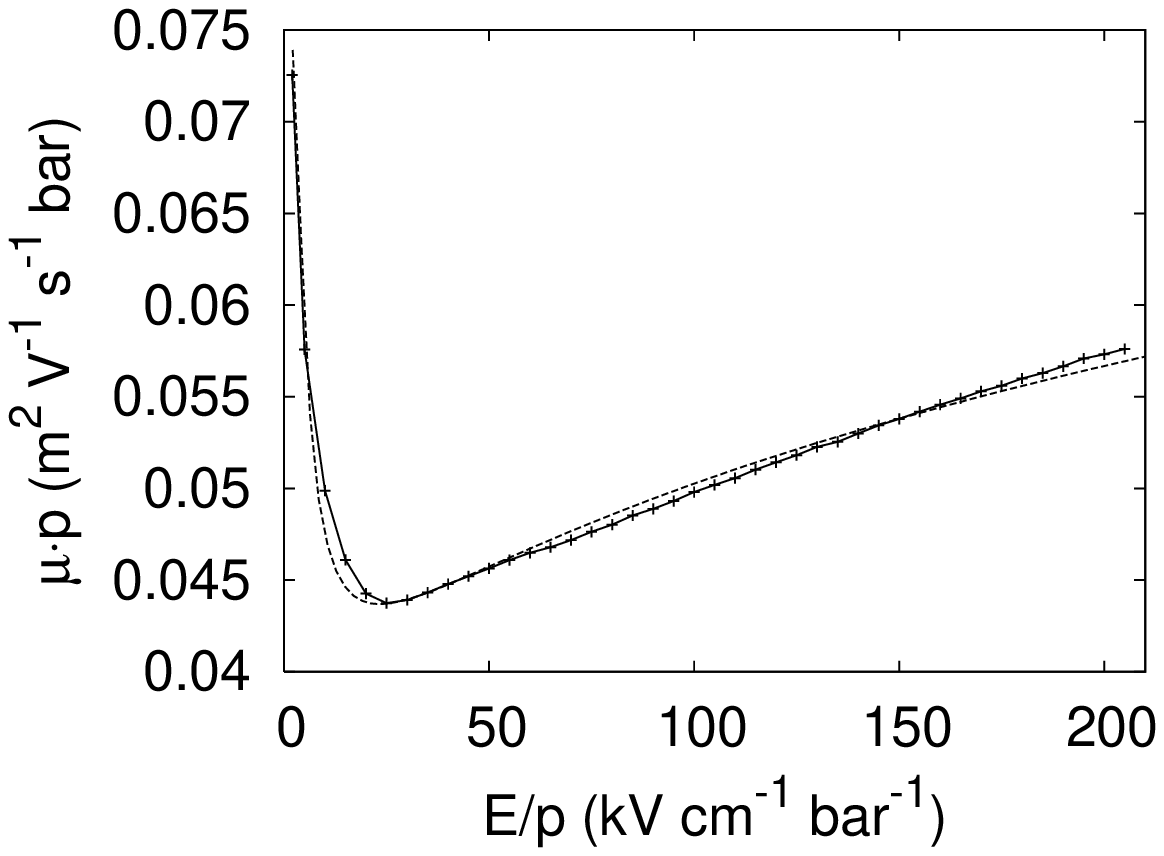}}
    \subfigure[ ~Ionization rate $\alpha(E)$]{
           \label{fig:coe_al}
           \includegraphics[width=.47\textwidth]{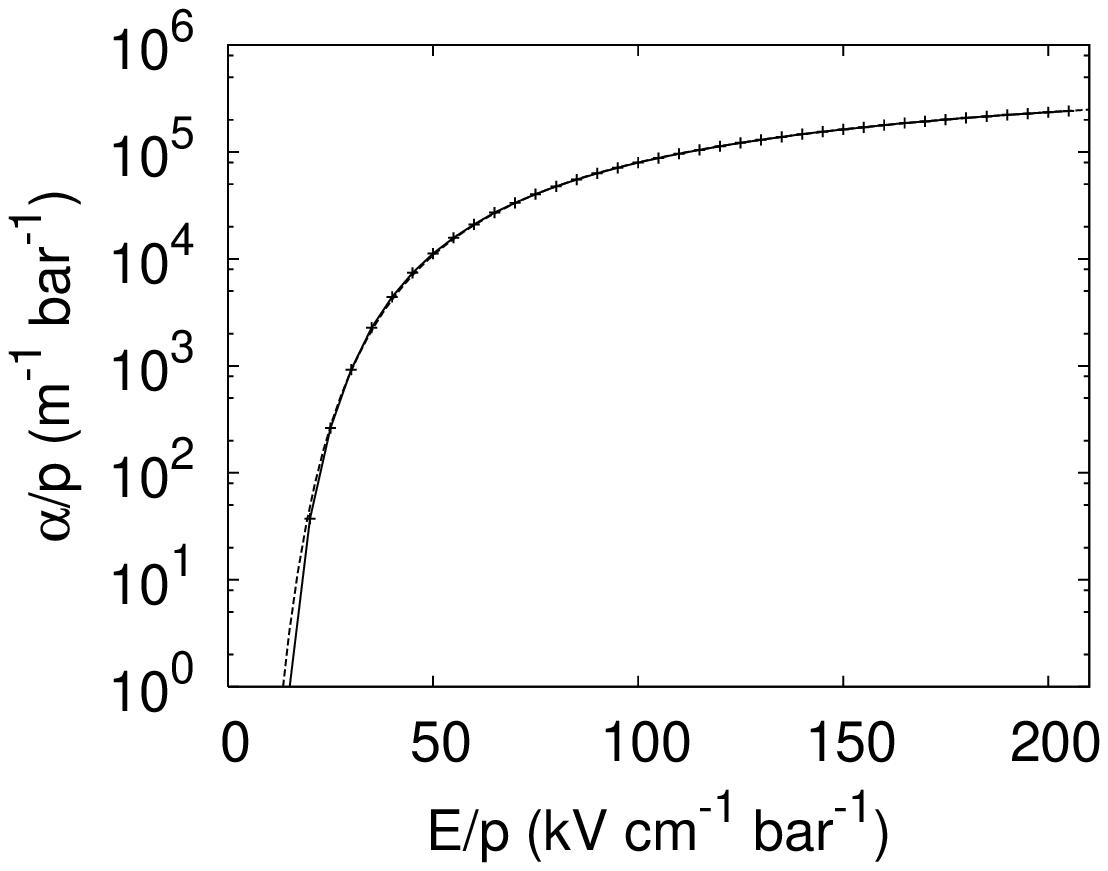}}
    \subfigure[ ~Transversal diffusion $D_T(E)$]{
          \label{fig:coe_dr}
          \includegraphics[width=.47\textwidth]{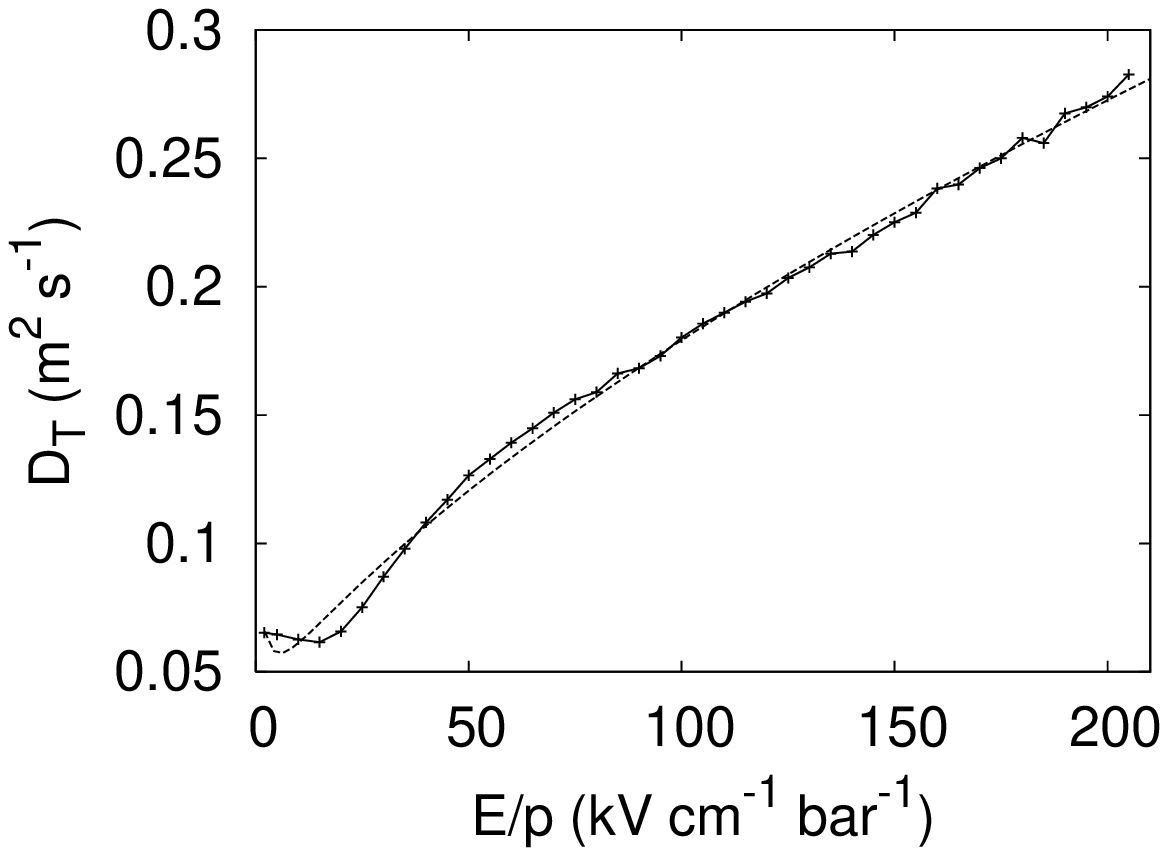}}
    \subfigure[ ~Longitudinal diffusion $D_L(E)$]{
          \label{fig:coe_dl}
          \includegraphics[width=.47\textwidth]{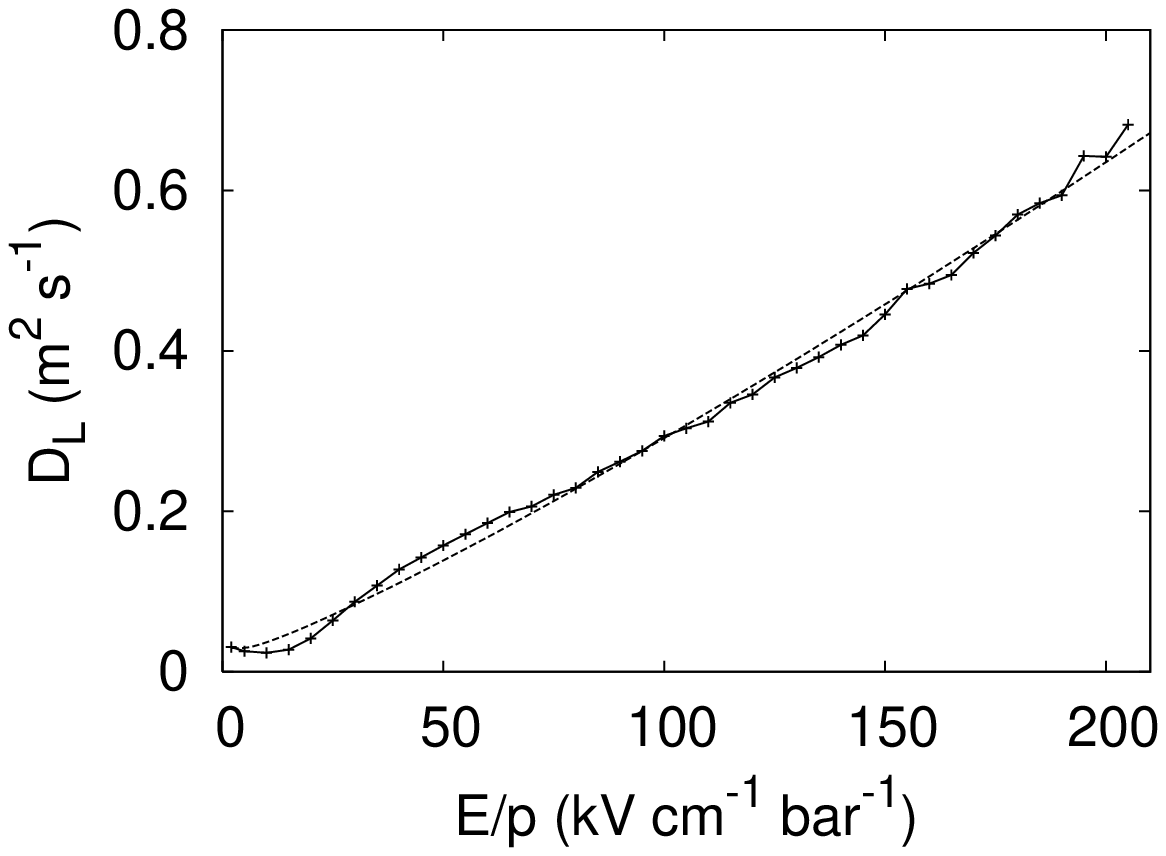}}
    \subfigure[ ~Average energy $\varepsilon(E)$]{
          \label{fig:coe_en}
          \includegraphics[width=.47\textwidth]{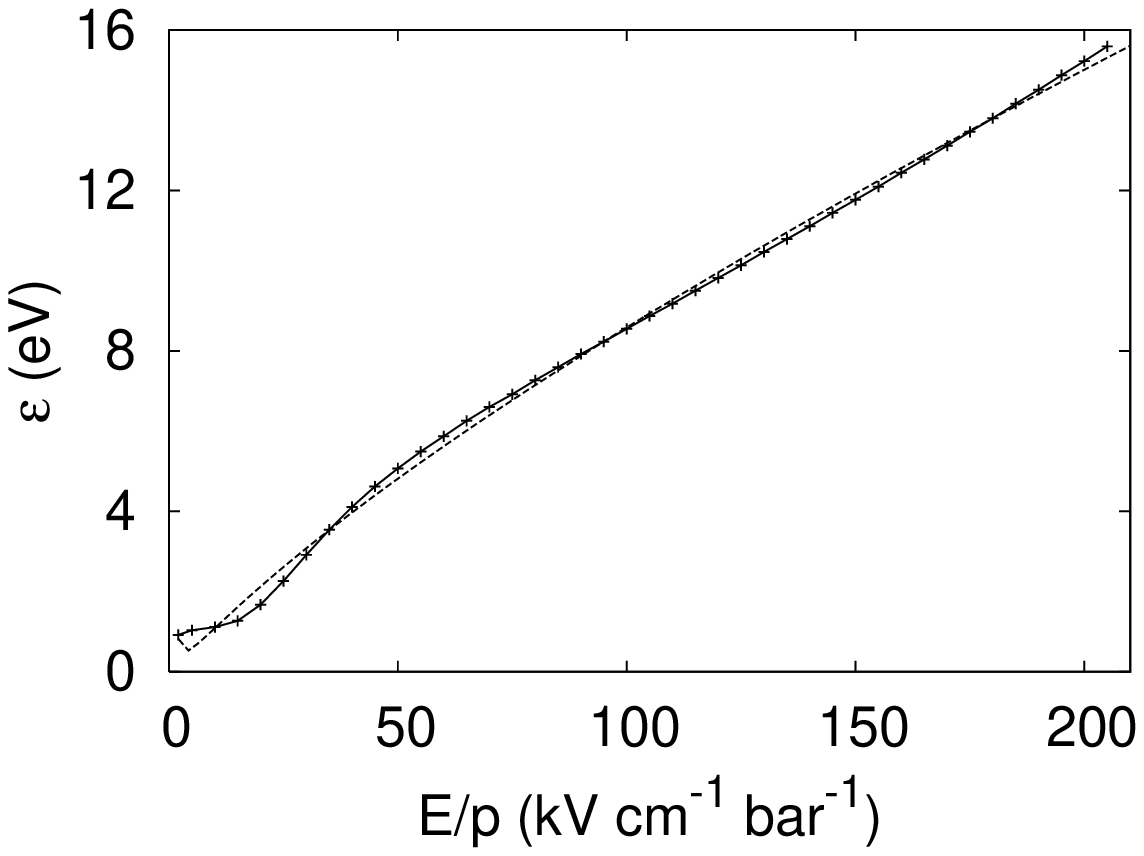}}
 \centering
 \caption{Electron mobility, diffusion rates, ionization rate and average electron energy in nitrogen. Plotted are the reduced coefficients $\mu p$, $\alpha /p$, $D_ T$, $D_L$ and $\varepsilon$ as a function of reduced field $E/p$ at room temperature. Our units are related to other commonly used units like $1$ kV cm$^{-1}$ bar$^{-1}$ = $1.316$ V cm$^{-1}$ torr$^{-1}$ = $0.424$ Townsend at T=$300$ K. 
 }
  \label{fig:trancoe}
\end{figure*}

The numerical results for $\mu$, $D_T$, $D_L$, $\alpha$ and $\epsilon$ for different electric fields $E=|{\bf E}|$
are presented in Fig.~\ref{fig:trancoe} together with empirical fit formulas. These formulas are
\begin{eqnarray}\label{equ:empirical1}
\mu(E)    & = & \frac{\rm m^2}{\rm V \cdot s} \cdot \exp\left[-4.02 + 0.21\ln \frac{ E}{\rm kV/cm }\right. \nn \\
&& \left.+ \frac{5.44\;\rm kV/cm }E -\;\left(\frac{2.42\;\rm kV/cm}{E}\right)^2\right]  
\end{eqnarray}
\begin{eqnarray}\label{equ:empirical2}
\alpha(E) & = & \frac{1}{\rm m}\cdot\exp\left[\;12.5 + 0.16\ln \frac{ E}{\rm kV/cm } \right. \nn \\
&&  \left.- \frac{200\;\rm kV/cm } E+ \;\left(\frac{19.2\;\rm kV/cm }{E}\right)^2\right] 
\end{eqnarray}
\begin{eqnarray}\label{equ:empirical3}
 D_T(E)   & = & \frac{\rm m^2}{\rm s}\cdot\exp\left[-4.71 + 0.64\ln \frac{ E}{\rm kV/cm } \right. \nn \\
&& \left.+ \frac{4.80\;\rm kV/cm} E-\;\left(\frac{1.84\;\rm kV/cm} {E}\right)^2\right] 
\end{eqnarray}
\begin{eqnarray}\label{equ:empirical4}
 D_L(E)   & = & \frac{\rm m^2}{\rm s}\cdot\exp\left[-6.75 + 1.18\ln \frac{ E}{\rm kV/cm}   \right.\nn \\
&&  \left.+ \frac{7.89\;\rm kV/cm} E-\;\left(\frac{2.49\;\rm kV/cm}{E}\right)^2\right] 
\end{eqnarray}
\begin{eqnarray}\label{equ:empirical5}
\varepsilon(E) & = & {\rm eV} \cdot \exp\left[-1.37 + 0.78\ln \frac{ E}{\rm kV/cm} \right. \nn \\
&&  \left. -\frac{4.44\;\rm kV/cm} E+ \;\left(\frac{3.46\;\rm kV/cm}{E}\right)^2\right],
\end{eqnarray}
under normal conditions. 

These coefficients will be used in the fluid model (\ref{fluid1})--(\ref{eq:np}) to reach optimal agreement between particle and fluid model.

\section{Simulations of planar fronts}\label{sec:result}

\subsection{Concepts and set-up of planar ionization fronts}\label{sec:con_pf}

\subsubsection{The role of planar fronts in the inner analysis of the streamer structure}\label{sec:model_pf}

Fluid model simulations of streamers within the past 20 years (see e.g.,~\cite{Dha1987,Vit1994,Mon2006:3,Luq2007}) have shown
that the streamer head is surrounded and preceded by an ionization front that propagates into the non-ionized gas.
Within the ionization front, ionization grows until space charge effects set in. The formed space charge layer is much thinner
than the radius of the streamer, it leads to a screening of the electric field in the interior of the streamer head and to
a field enhancement ahead of it. Therefore the field dependent ionization reaction coefficient $\alpha(E)$ is enhanced ahead
of the space charge layer and suppressed behind it. The space charge layer around the streamer is shown in
Fig.~\ref{fig:planarfront}, for a further discussion of the three-dimensional structure and growth of streamers,
we refer to the literature, see, e.g., \cite{Ebe2006}.

It is clear that the full configuration of the electric field can only be analyzed within a two- or
three-dimensional setting. On the other hand, within the inner structure of ionization front and space charge
layer, the electric field does not deviate much from a planar configuration. To analyze the processes within the
ionization front in detail, it is therefore advisable to study the inner structure of a planar front. This will be
done here. The results can be put in further physical context through a separate analysis on the inner and the
outer scale of the structure as commonly done in hydrodynamic boundary layer analysis, reaction-diffusion systems
etc.~\cite{Ben1978, Fif1988, Ebe2000:1}.

\subsubsection{Construction of planar fronts in the particle model}

The construction of a planar front is straight forward in the fluid model: gradients $\nabla$ are simply
evaluated in one spatial direction. We choose this direction to be the $z$ direction. In the particle model, on
the other hand, electrons move in all three spatial dimensions. Therefore a three-dimensional setting has to be
retained. An essentially one-dimensional setting is achieved by considering only a small transversal area $A$ of
the front and by imposing periodic boundary conditions at the lateral boundaries. Furthermore, the electric field
is calculated only in the forward direction $z$ through \be \label{eq:1dPoisson}
E_z(z,t)=E_z(z_0,t)+\int_{z_0}^zdz' \int_A \frac{dx\;dy}A~\frac{{\rm e}(n_p-n_e)(x,y,z',t)}{\epsilon_0} \ee which
is the one-dimensional version of (\ref{eq:Poisson}). This means that fluctuations of the transversal field due to
density fluctuations in the transversal direction are not included. In fact, the numerical implementation of
\eqref{eq:1dPoisson} is performed on a grid in the forward direction only as discussed in section~\ref{sec:num-part}.

The density fluctuations projected onto the forward direction depend on the transversal area $A$ over which the
averages are taken. When $A$ increases, the total number of electrons in the simulation increases proportionally
to $A$, while the relative density fluctuations decrease like $1/\sqrt{A}$. Therefore some intermediate value of the
area $A$ has to be chosen: On the one hand, there should be a sufficient number of electrons to reach a satisfactory
statistics, but on the other hand, there shouldn't be so many electrons that the computer runs out of memory within
the time interval of interest. In the simulation, we use a small $A$ for high electric field and a large $A$
for low electric field. For example, we choose the transversal averaging area as $A=6\; \Delta \ell \times 6\; \Delta \ell$ for $-100$ kV/cm,
but for $-50$ kV/cm, the transversal averaging area is $A=20\; \Delta \ell \times 20\; \Delta \ell$, here $\Delta \ell=2.3\;\mu{\rm m}$.

In the $z$ direction, the system length is $500\; \Delta \ell$ which allows the front to propagate freely for all runs
reported in this paper. The electric field in the non-ionized region at large $z$ is specified by ${\bf E} =
E^+\;\hat{\bf z}$, where $E^+<0$ and $\hat{\bf z}$ is the unit vector in the $z$ direction.

In the simulations, two different initial conditions are used. In the first, $N_e$ electrons are evenly
distributed in a thin layer of area $A$ with an extension of $19.5\;\Delta \ell < z < 20.5\;\Delta \ell$ in the field direction.
Choosing
\begin{equation}
N_e/A=|E^+| \cdot \epsilon_0 / {\rm e} \label{eq:initialcharge};
\end{equation}
the field behind the layer is screened according to (\ref{eq:1dPoisson}). Another choice is to
begin with a few seed electrons which will create an ionization avalanche and form a charge layer later.

\subsection{Planar fronts in the particle model}

\subsubsection{Qualitative discussion of typical results}

\begin{figure*}
    \centering
      \includegraphics[width=0.95\textwidth]{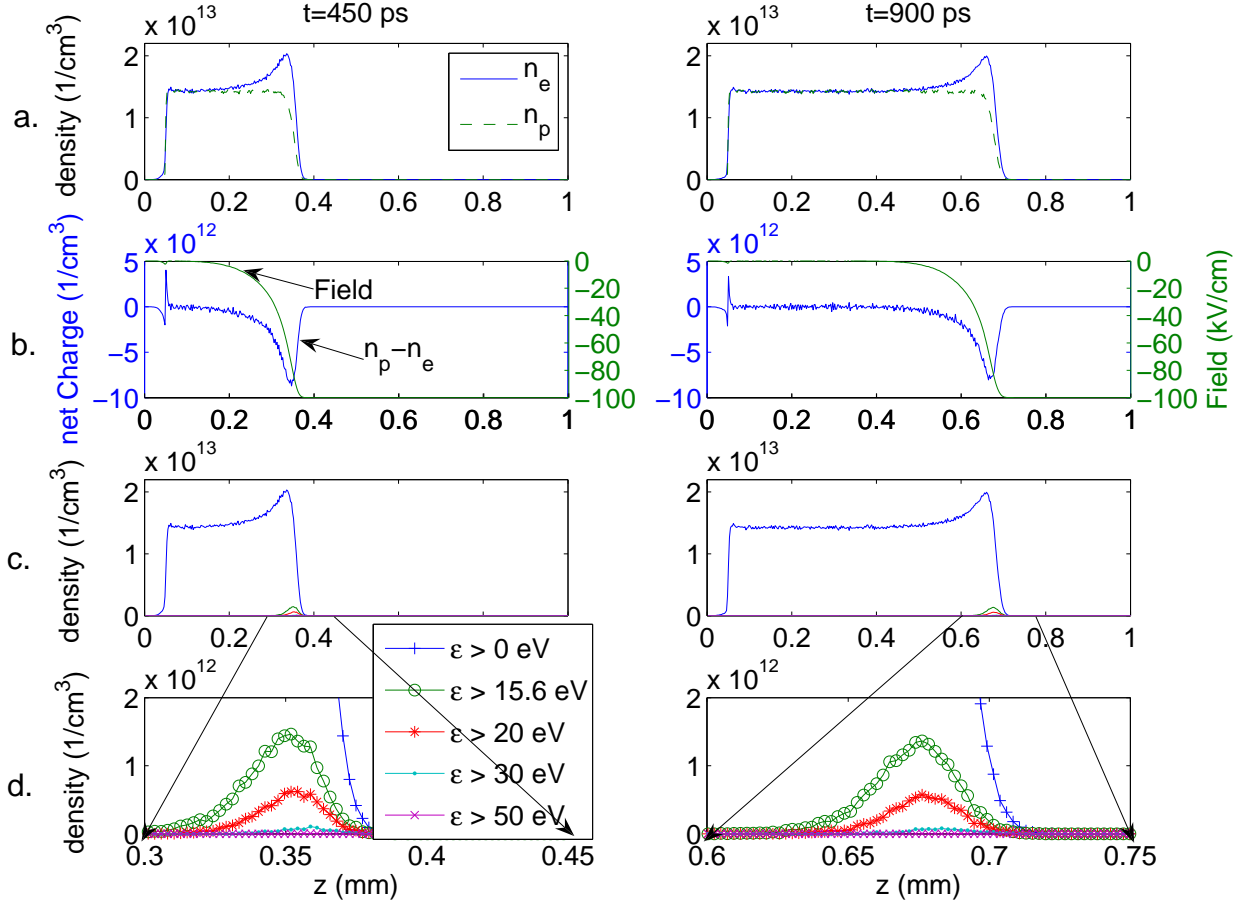}
  \centering
\caption{\label{fig:pfdensity} Spatial profile of a particle front within a field of $-100$ kV/cm at time
 $t_1=450$~ps (left) and $t_2=900$~ps (right). (a) Electron density (solid) and ion density (dashed),
 (b) electric field (dashed) and net charge (solid), (c) density of electrons with energy above 0 eV, 15.6 eV,
 20 eV, 30 eV and 50 eV, (d) zoom into the front region with high electron energies, same quantities as in (c).}
\end{figure*}

We first present results in a field of $E^+=-100$~kV/cm. The initial condition is a thin electron layer with total
electron number $N_e$ as in (\ref{eq:initialcharge}), screening the electric field behind the layer.
Fig.~\ref{fig:pfdensity} presents the evolution at times $t_1=450$ ps (left) and $t_2=900$ ps (right). Panel (a)
shows the density distribution of the electrons (solid) and the ions (dashed). Panel (b) shows the net negative
charge distribution (solid) and the electric field (dashed). Panel (c) shows the total charge density of electrons
and the charge density of electrons with an energy higher than 0, 15.6, 20, 30, and 50 eV, where 15.6 eV is the
ionization energy. Panel (d) zooms into panel (c), both in space and in charge densities.

The figure shows an ionization front propagating to the right; up to fluctuations, the spatial profiles are
essentially unchanged, therefore the front velocity $v$ is essentially constant as well. The front carries an
overshoot of electrons, generating a thin space charge layer that screens the electric field behind the front.
In this screened interior region, the electron and ion density reach an equal constant density $n_e^-=n_p^-$. (Upper indices ``$\pm$'' indicate quantities before ``$+$'' and behind ``$-$'' the ionization front.)
The qualitative features of the front are the same as those in the fluid model~\cite{Ebe1997}.

Electrons with energies above the ionization threshold of 15.6~eV are shown in the lower two panels in
Fig.~\ref{fig:pfdensity}; they exist essentially only in the high field region. Electrons with energies above
30~eV are so rare that they are hardly seen even on the scale of panel (d). Electrons with energy above 50~eV
exist, but cannot be distinguished within this plot. The profiles of high energy electrons also move with the
whole front without change of shape, up to fluctuations.

Following the track of single electrons of high energy, we found that they gain and loose energy in few collisions
within a few ps and do not run away. In that sense they are in a fast dynamic equilibrium with the electrons
at lower energies. This observation agrees with the fast relaxation of 50 eV electrons travelling in the forward
direction whose energy relaxation is shown in figure~\ref{fig:en_relax}.

\subsubsection{Quantitative results in different fields}

The ionization front in a given field $E^+$ is characterized by a velocity $v$, a degree of ionization $n_e^-=n_p^-$
behind the front and an electron energy distribution in the high field region. We now present these quantities in detail.

We define the front position as the position of the maximal electron density.
Table~\ref{tab:particl_pf} summarizes our numerical results on the front velocity $v$ as well as on the saturated
electron density $n_e^-$ behind the front as a function of the electric field $E^+$ immediately ahead of the front.

\begin{table}
\begin{tabular}{c|c|c}
\hline
$E^+$ & $v$ & $n_e^-$ \\
(kV/cm) & (m/s) & (1/ cm$^3$) \\
\hline
 $ 50  $  &  ${\quad}$ $(2.773 \pm  0.007) \cdot 10^5$ ${\quad}$ & ${\quad}$ $( 5.923 \pm 0.031 )\cdot 10^{11} $ ${\quad}$ \\
 $ 75  $  &  $(4.845 \pm  0.023) \cdot 10^5$ &  $ ( 4.372 \pm 0.011 )\cdot 10^{12}  $ \\
 $ 100 $  &  $(7.258 \pm  0.062) \cdot 10^5$ &  $ ( 1.422 \pm 0.003 )\cdot 10^{13} $ \\
 $ 125 $  &  $(1.012 \pm  0.010) \cdot 10^6$ &  $ ( 3.233 \pm 0.007 )\cdot 10^{13} $ \\
 $ 150 $  &  $(1.365 \pm  0.008) \cdot 10^6$ &  $ ( 6.014 \pm 0.006 )\cdot 10^{13}  $ \\
 $ 175 $  &  $(1.745 \pm  0.027) \cdot 10^6$ &  $ ( 9.875 \pm 0.020 )\cdot 10^{13}  $ \\
 $ 200 $  &  $(2.262 \pm  0.063) \cdot 10^6$ &  $ ( 1.486 \pm 0.004 )\cdot 10^{14}  $ \\
\hline
\end{tabular}
\caption{Numerical results on planar fronts in the particle model: front velocity $v$ and ionization level $n_e^-$
behind the front as a function of the electric field $E^+$ ahead of the front.} \label{tab:particl_pf}
\end{table}

\subsection{Planar fronts in the fluid model}

For planar fronts in the fluid model, there are not only numerical, but also analytical results, both agree within the numerical accuracy. First, the velocity of the front in a field $E^+$ is
given by \be \label{v} v^*=\mu(E^+)\;|E^+|+2\sqrt{D_L(E^+) \; \mu(E^+) \; |E^+| \; \alpha(E^+)}, \ee according
to~\cite{Ebe1997,Lag1994} with a slight generalization along the general lines discussed in~\cite{Ebe2000}. Note
that for the initial conditions treated in this paper, the velocity $v^*$ is approached from below after an
algebraically slow relaxation~\cite{Ebe2000}.

The electron and ion densities decay exponentially like $e^{-z/\ell^*}$ in the leading edge of the front where the
electric field is approximately $E^+$~\cite{Ebe1997,Ebe2000}. The decay length is \be \label{ell}
\ell^*=\sqrt{\frac{D_L(E^+)}{\mu(E^+)\;|E^+|\;\alpha(E^+)}}. \ee These analytical results are summarized in
Table~\ref{tab:analy_pf}.

\begin{table}
\begin{tabular}{c|c|c}
\hline
$E^+$ & $v^*$ & $\ell^*$\\
(kV/cm) & (m/s) & (m) \\
\hline
 50   & ${\quad}$ $2.68\; \cdot 10^5 $ ${\quad}$ & ${\quad}$ $\;7.83\; \cdot 10^{-6} $ ${\quad}$ \\
 75   & $4.70\; \cdot 10^5 $ & $\;3.91\; \cdot 10^{-6} $ \\
 100  & $7.14\; \cdot 10^5 $ & $\;2.72\; \cdot 10^{-6} $ \\
 125  & $9.88\; \cdot 10^5 $ & $\;2.15\; \cdot 10^{-6} $ \\
 150  & $1.29\; \cdot 10^6 $ & $\;1.84\; \cdot 10^{-6} $ \\
 175  & $1.63\; \cdot 10^6 $ & $\;1.66\; \cdot 10^{-6} $ \\
 200  & $1.98\; \cdot 10^6 $ & $\;1.54\; \cdot 10^{-6} $ \\
\hline
\end{tabular}
\caption{Exact analytical results for planar fronts in the fluid model, evaluated with the transport coefficients from
\eqref{equ:empirical1}--\eqref{equ:empiricalL}: front velocity $v^*$ and electron density decay length $\ell^*$
as a function of the electric field $E^+$.} \label{tab:analy_pf}
\end{table}

For the degree of ionization behind the front $n_e^-$, there is no closed analytical solution. The ionization behind
the front can be derived numerically.
Furthermore, in the appendix, we derive an analytical upper bound for this quantity, namely \be \label{ion-bound}
n_e^-\lessapprox n_{e,bound}^-=\frac{\epsilon_0}{\rm e} \int_0^{|E^+|}\alpha(e)\;de. \ee Numerical result and
analytical bound are summarized in table~\ref{tab:numeri_pf}.

\begin{table}
\begin{tabular}{c|c|c}
\hline
$E^+$ & $n_e^-$ &  $n_{e,bound}^-$ \\
(kV/cm) & (1/ cm$^3$) & (1/ cm$^3$) \\
\hline
 50  & ${\quad}$ $5.43\; \cdot 10^{11} $ ${\quad}$ & ${\quad}$  $5.80\; \cdot 10^{11} $ ${\quad}$ \\
 75  &  $3.83 \; \cdot 10^{12} $ &  $3.98 \; \cdot 10^{12} $ \\
 100 &  $1.17 \; \cdot 10^{13} $ &  $1.22 \; \cdot 10^{13} $ \\
 125 &  $2.49 \; \cdot 10^{13} $ &  $2.61 \; \cdot 10^{13} $ \\
 150 &  $4.35 \; \cdot 10^{13} $ &  $4.58 \; \cdot 10^{13} $ \\
 175 &  $6.70 \; \cdot 10^{13} $ &  $7.09 \; \cdot 10^{13} $ \\
 200 &  $9.50 \; \cdot 10^{13} $ &  $10.1 \; \cdot 10^{13} $ \\
\hline
\end{tabular}
\caption{The degree of ionization $n_e^-$ behind the front in the fluid model as a function of field $E^+$:
numerical result $n_e^-$ and analytical upper bound $n_{e,bound}^-$ as derived in the appendix.} \label{tab:numeri_pf}
\end{table}

\subsection{Comparison of planar fronts in particle and fluid model}

\subsubsection{Detailed investigation in a field of $-100$ {\rm kV/cm}}

Now the stage is set to compare planar fronts in the particle model to those in the fluid model,
The comparison is first done in detail
for a planar front propagating into a field of $-100$~kV/cm. Both fluid and particle simulations are carried out
in the same setup starting from the same initial conditions, i.e., from an electrically neutral group of 100
electrons and ions evenly distributed within the thin layer $19.5\;\Delta \ell < z < 20.5 \; \Delta \ell$ and within the transversal
area $A=6\; \Delta \ell \times 6\;\Delta \ell$.

\begin{figure}
  \begin{center}
      \includegraphics[width=0.5\textwidth]{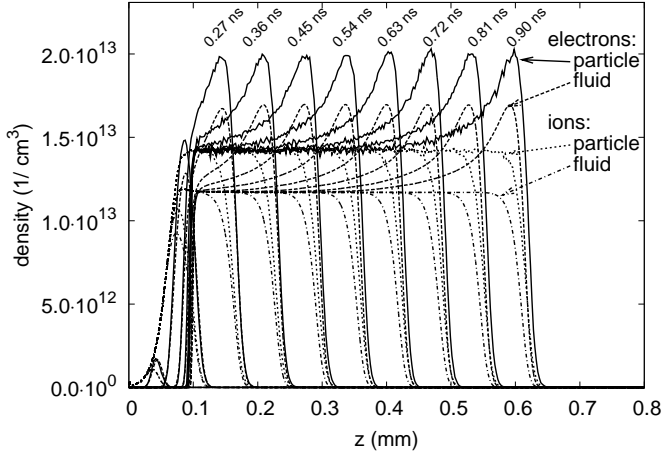}
 \end{center}
 \caption{\label{fig:1dfpcom} Temporal evolution of the electron and ion densities in a planar front in a field
 of 100~kV/cm. Shown are the spatial profiles of electron and ion densities derived with the particle or the fluid
 model at time steps $t = 0$~ns, 0.09~ns, 0.18~ns, 0.27~ns, $\ldots$, 0.9~ns (solid lines: $n_{e,part}$, dashed:
 $n_{e,fluid}$, dotted: $n_{p,part}$, dot-dashed: $n_{p,fluid}$).}
\end{figure}

Fig.~\ref{fig:1dfpcom} shows the temporal evolution of the planar front. We compare the spatial profile of the
electron density (solid line) and ion density (dotted) in a particle simulation with the electron density (dashed)
and ion density (dot-dashed) in a fluid simulation. Two features are clearly visible: First, the particle and the
fluid front move with approximately the same velocity and the same density profile in the leading edge of the
front where the electric field does  not vary yet. Second, the maximal electron density in the front and the
saturation level of the ionization behind the front in the particle model are about 20~\% higher than in the fluid
model. These results of visual inspection agree with those of tables I and III for a field of $-100$~kV/cm.

As we have excluded other reasons of the discrepancy like numerical discretization errors or inconsistent transport
and reaction coefficients, deviations must be due to the approximations in the fluid model, and a closer inspection
shows that we should focus on the electron energies.

\begin{figure}
     \begin{center}
          \includegraphics[width=.5\textwidth]{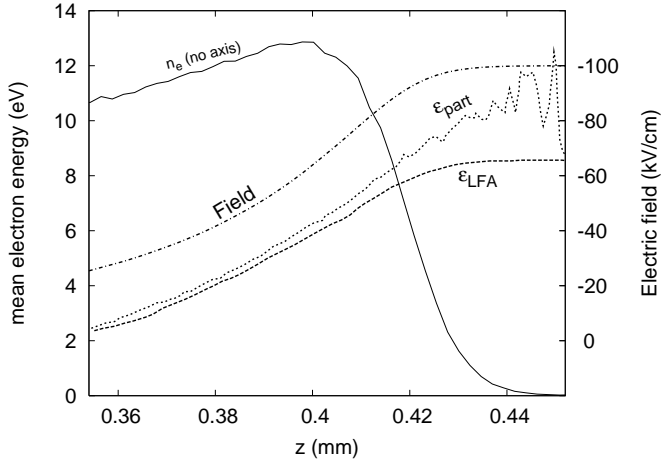}
 \caption{\label{fig:lfazoom} Zoom into the particle ionization front of Fig.~\ref{fig:1dfpcom} in a field of
 $-100$~kV/cm at time 0.63~ns: shown are electron density distribution $n_e$ (solid line), local average electron
 energy $\varepsilon_{\rm part}$(dotted line), local average electron energy $\varepsilon_{\rm LFA}$ according to
 the local field approximation (dashed line), and electric field strength $E$ (dot-dashed line).}
\end{center}
\end{figure}

Fig.~\ref{fig:lfazoom} zooms into the ionization front shown in Fig.~\ref{fig:1dfpcom} at time $t=0.63$~ns. Here
we show the electron density (solid line) and electric field (dot-dashed line) in the particle model. Furthermore
the local mean energy of the electrons in the particle model is indicated with a dotted line. Finally, the mean
electron energy according to the local field approximation $\epsilon(E)$ is derived from the local field $E$ and
equation (\ref{equ:empiricalL}) for $\epsilon(E)$; it is indicated with a dashed line. It can be seen that the
average electron energy nicely follows the local field approximation in the interior of the ionized region while
it is considerably higher in the region where the electric field is large and the electron density decreases
rapidly.

\begin{figure}
  \begin{center}
      \includegraphics[width=0.5\textwidth]{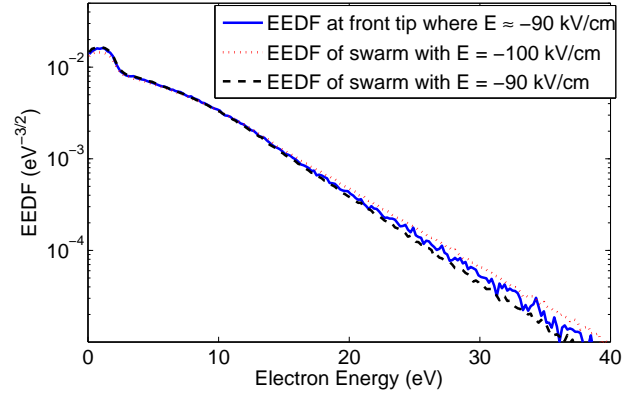}
 \caption{\label{fig:eedfcomparision} The electron energy distribution function is measured as $P(\varepsilon) / \sqrt{\varepsilon}$, where $P(\varepsilon)$ is the probability of electron energy $\varepsilon$. Here we show the electron energy distribution at the front region of Fig.~\ref{fig:lfazoom}
 where $E \approx -90$~kV/cm (solid line) and in the swarm experiments in constant fields of $-90$~kV/cm (dashed
 line) and $-100$~kV/cm (dotted line).}
\end{center}
\end{figure}

This deviation is analyzed in more detail in Fig.~\ref{fig:eedfcomparision} where not only the mean energy, but
the full electron energy distribution is shown. This is done for the particular spatial region of the front where
the electric field has decreased by 10 \% to a value of $-90$~kV/cm. More precisely, electrons are collected from
the first cell where $|E|>90$~kV/cm, searching with increasing $z$. The average field in these cells is about
$E=-91.47$~kV/cm. To reach a satisfactory statistics, the electrons are collected from 10 different instants of
time with a temporal distance of 30 ps between the two consecutive sampling times to ensure statistical
independence. While these data are plotted as a solid line, for comparison the electron energy distributions in the
swarm experiments in a constant field of $-90$ and $-100$ kV/cm are plotted as a dashed or dotted line,
respectively. Analyzing the electron energy distribution at low energies $\varepsilon < 10$ eV, the ionization front
at a local field around $-90$kV/cm and the swarm experiment in a constant field of $-90$~kV/cm are quite similar while
the distribution for a swarm in a field of $-100$ kV/cm is clearly lower. On the other hand, for electron energies
above 20 eV, the energy distribution in the ionization front at a local field of $-90$ kV/cm actually lies closer
to the distribution of the swarm at $-100$ kV/cm than to that at $-90$ kV/cm.
This observation not only confirms that the average electron energy in the leading edge of the ionization front
is higher than in the local field approximation, as is consistent with Fig.~\ref{fig:lfazoom}, but it indicates
that the higher mean energies correspond to a higher population of electron energy states above the ionization
threshold. We therefore expect that also the ionization rates are higher in the particle model than in the local
field approximation.

\subsubsection{Results for other fields}

\begin{figure}
\begin{center}
    \includegraphics[width=.5\textwidth]{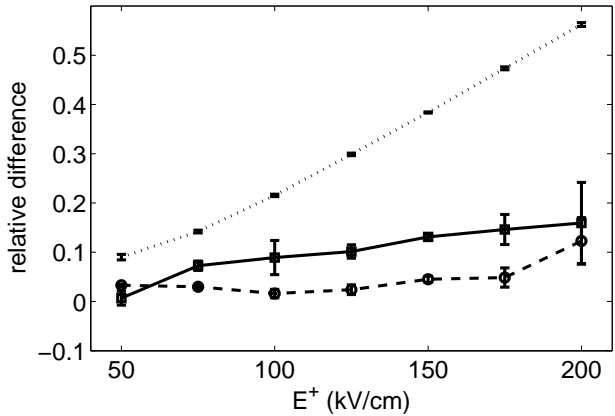}
    \caption{\label{fig:discrepancy} Relative difference of ionization level behind the front \eqref{n-rel}
    (dotted, above), front velocity \eqref{v-rel} (dashed, below), and mean electron energy in the leading edge \eqref{eps-rel} (solid, middle) between particle and
    fluid model as a function of the electric field $E^+$ ahead of the front.
    }
\end{center}
\end{figure}

Having analyzed the front propagating into a field of $E^+=-100$~kV/cm in detail, we now summarize equivalent
results for fields ranging from $-50$ to $-200$~kV/cm in Fig.~\ref{fig:discrepancy}. The figure shows the relative
difference \be \label{n-rel}\frac{n^-_{e,\;\rm part.}-n^-_{e,\; \rm fluid}}{n^-_{e,\;\rm fluid}} \ee of the saturated
electron density behind the front between the particle and the fluid model, the relative difference \be \label{v-rel}\frac{v_{\rm part}-v_{\rm fluid}}{v_{\rm fluid}} \ee of the front velocity and the relative difference
\be \label{eps-rel} \frac{\langle \varepsilon_{\rm part}\rangle -\langle \varepsilon_{\rm LFA} \rangle}
{\langle \varepsilon_{\rm LFA}\rangle} \ee of the mean electron energy between the particle model and  the local
field approximation at a point in the ionization front where electron densities are low and the electric field is
just slightly screened to the value $E=0.98\; E^+$. All differences increase with increasing field.

\subsection{Interpretation of results}

\subsubsection{Electric field gradient, electron motion in the front and electron energy overshoot}

A few authors have dealt with particle discharge models with strong gradients both in the electric field and in the
particle density. They analyzed deviations between particle models and the fluid model in local approximation by means
of the Boltzmann equation, as treated in section \ref{sec:Boltz}. They analyzed the case of stationary gradients of
the electric field~\cite{sat1985,Ale1996} and the case of positive streamer ionization fronts~\cite{Nai1997}.
They suggested corrections to the fluid model both due to field gradients and to density gradients.

The situation in negative streamer ionization fronts is somewhat different: 
a negative ionization front in nitrogen moves approximately with the electron drift velocity determined by the electric
field in the leading edge of the ionization front. This means that the electrons in this leading edge region on
average move within a stationary electric field. More precisely, the front velocity $v^*(E^+)$ from (\ref{v}) is slightly
larger than the electron drift velocity $\mu(E^+)\;|E^+|$ in the electric field ahead of the front. This is because
the front is not carried by electron drift only, but also by diffusion and creation of new free electrons. In a coordinate system
moving with the ionization front, the existing electrons therefore on average move backwards. This motion is the faster,
the further they are behind.

Fig.~\ref{fig:eedfcomparision} shows that the high energy tail of the electron energy distribution in the
ionization front at the position where the field is $-91.47$~kV/cm, is closer to the swarm experiment at
$-100$~kV/cm than to that at $-90$~kV/cm. One could interpret these results by assuming that the electrons have
gained their energy distribution in a field of close to $-100$~kV/cm and then are transported backwards to where the
field is only $-90$~kV/cm.

However, Fig.~\ref{fig:lfazoom} shows that this interpretation cannot be true. If the electron energy distribution
would first ``equilibrate'' to a field of $-100$~kV/cm and then partially be transported backwards, then the mean
electron energy $\langle \epsilon_{\rm part}\rangle$ would everywhere be below the mean electron energy $\langle
\epsilon_{\rm LFA} \rangle$ at $-100$~kV/cm in local field approximation. But clearly there is an electron energy
overshoot in the leading edge of the ionization front. So the most prominent deviation from the fluid model visible
in Fig.~\ref{fig:lfazoom} occurs in the region were the electric field is almost constant; furthermore, if it would
be a consequence of the local field approximation, the deviation would have the opposite sign.

\subsubsection{Density gradient and relation between front and swarm experiments}

We conclude that the electron energy overshoot in the leading edge of the ionization front has to be due to the
electron density gradient rather than to the field gradient. The effect of density gradients can be tested in
swarm experiments (cf.~section~\ref{swarm}) in a constant electric field as well. It turns out that such a test
goes beyond qualitative results and actually allows for a quantitative comparison of fronts and swarms as will
be explained below.

The key to the quantitative comparison is based on two closely related facts: $(i)$ the density profiles
in the leading edge of a swarm or avalanche in a constant field and in the leading edge of an ionization front
penetrating the same field are both given by \be \label{eq:prof} n_e \sim e^{-z/\ell^*}, \ee where $\ell^*$
is given by \eqref{ell} and in table II. $(ii)$ Also the velocities of swarm and front have the same value $v^*$
from \eqref{v} and table II.

This relation between velocities and decay rates of swarms and fronts holds generally for any so-called
pulled front whose velocity is determined in the linearly unstable region ahead of the front, as is discussed
in a general setting in \cite{Ebe2000}, specifically in section 2.5.1. The statement can be verified
on the explicit form of the Gaussian profile \eqref{Gauss}: in the coordinate \be \label{eq:xi} \xi=z-z_0-v^*t \ee
moving with velocity $v^*$, the Gaussian swarm distribution (\ref{Gauss}) can be written as \be n_e\propto
e^{\;\mu|E|\alpha t}\;\frac{e^{-(z-z_0-\mu Et)^2/(4D_Lt)}}{\sqrt{4\pi D_L t}} =
e^{-\xi/\ell^*}\;\frac{e^{-\xi^2/(4D_Lt)}}{\sqrt{4\pi D_L t}}. \ee

\begin{figure*}
     \centering
     \includegraphics[width=0.9\textwidth]{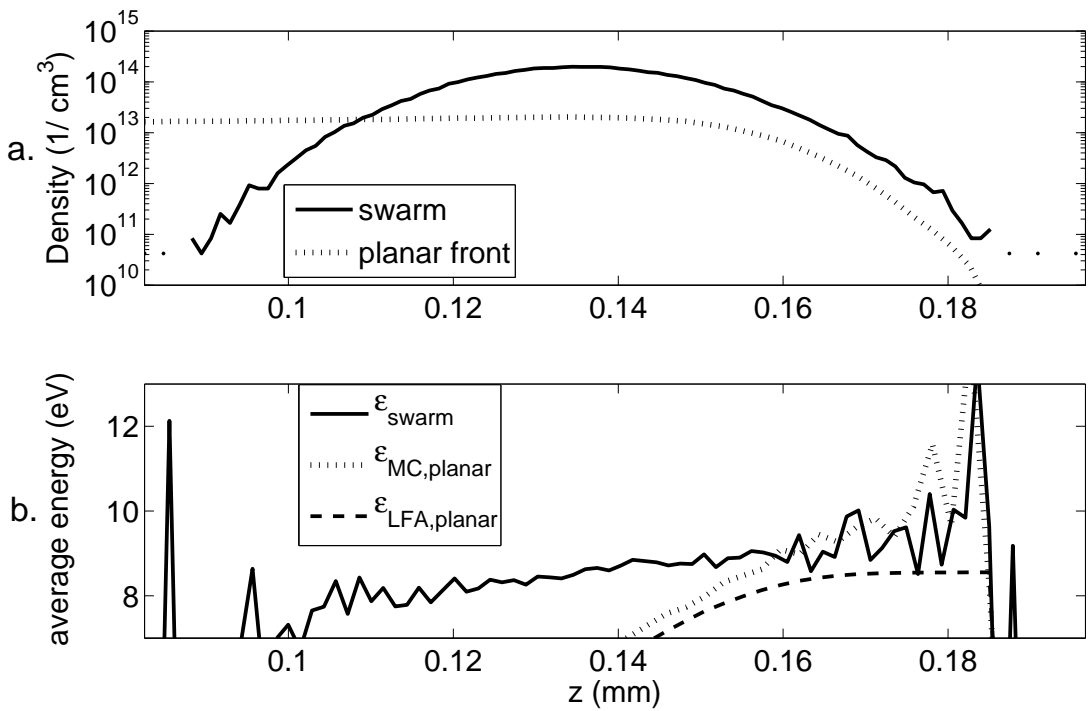}
 \centering
 \caption{Comparison of an electron swarm and a planar ionization front, both in the particle model and in a field
 of $-100$ kV/cm. (a) Electron density on a logarithmic scale for swarm (solid) and front (dotted) as a function of $z$.
 (b) Mean electron energy in swarm (solid) and front (dotted) and in the front assuming the local field approximation
 (dashed) as a function of $z$.}
 \label{fig:den_en}
\end{figure*}

We now test the above theoretical predictions on the particle model for a swarm and a planar front in a field of
$-100$ kV/cm. The results are shown in Fig.~\ref{fig:den_en}. It should be remarked that our electron swarm has
a Gaussian density distribution both in the longitudinal and in the transversal direction. To focus on the profile
in the longitudinal direction, the density in the swarm in Fig.~\ref{fig:den_en}.a is taken as the density on the
longitudinal axis. Fig.~\ref{fig:den_en}.a shows the electron density profile of the swarm decaying at both edges and
the profile of the planar front that grows and saturates to a constant level. As the densities are plotted on a
logarithmic scale, an exponential decay like in \eqref{eq:prof} amounts to a straight line with slope $-1/\ell^*$
in the plot. Despite density fluctuations and slow transients in the buildup of the profile \cite{Ebe2000},
Fig.~\ref{fig:den_en}.a indeed shows that swarm and front have a similar decay profile on the right hand side.

This sets the stage to compare now the electron energies of swarm and front in Fig~\ref{fig:den_en}.b.
The dotted and the dashed line reproduce lines from Fig.~\ref{fig:lfazoom} for the planar particle front,
they show the actual mean electron energy $\langle\varepsilon_{\rm part}\rangle$ (dotted) and the mean electron
energy according to the local field approximation $\langle\varepsilon_{\rm LFA}\rangle$ (dashed).
For the electron swarm, the mean electron energy along the swarm axis is indicated as a solid line in
Fig~\ref{fig:den_en}.b. It is remarkable that this energy within the swarm is not constant in space
though the electric field has a constant value. Rather the energy increases almost linearly along the axis.
While the average energy of the swarm is around 8.5~eV, producing the respective value for the local field
approximation $\varepsilon(E)$ in Fig.~\ref{fig:coe_en}, the average energy at the leading edge of the swarm
reaches values above 10~eV.

Now the leading edge regions of swarm and front need a closer inspection. The leading edge is defined as the region
where the electric field is (almost) constant and where the electron densities both approach the profile \eqref{eq:prof}.
It is in this region where the electrons attain the highest mean energies, and Fig~\ref{fig:den_en}.b shows that
the mean energy profiles of swarm and front in this region are virtually identical up to fluctuations.
We conclude that the high electron energies within the leading edge of the ionization front that are shown
in figures \ref{fig:lfazoom} and \ref{fig:eedfcomparision}, are due to the electron density gradient, and can
be studied in the leading edge of a swarm experiment as well.

\subsubsection{The ionization density behind the front}

The discussion above shows that within the electron density gradient, the fast electrons are on average a bit ahead
of the slower electrons as they have a higher average velocity in the forward direction. This is understandable since
on the one hand, each single fast electron looses its energy over a length of $\sim 1$ $\mu$m (which is the electron
drift velocity $\mu(E)E$ times the relaxation time $\sim 2$~ps from figure~\ref{fig:en_relax}), but on the other hand,
the gradient length $\ell^*\approx 2.7$ $\mu$m of the density profile is of the same order. Clearly, yhis effect
becomes the more pronounced, the higher the field.

Now the electron energies in the leading edge substantially exceed the local field approximation and indicate the presence
of a larger fraction of electrons with energies above the ionization threshold. This leads also to higher ionization rates
than estimated by the local field approximation $\alpha(E)$ \eqref{equ:empiricalL}. As the ionization level behind the front
$n_e^-=n_p^-$ is well approximated by the effective ionization rate $\alpha$ integrated over the fields within the front (\ref{ion-bound})
(recall also the argument in the appendix), it is clear that the ionization level behind the front is higher in the particle
model than in the fluid model.

\section{Conclusion}\label{sec:conclusion}

Negative streamer ionization fronts in pure nitrogen were investigated in planar approximation, both following
the kinetics of single electrons in a particle model, and in a fluid model in local field approximation.
As parameter functions for the fluid model were derived from swarm experiments in the particle model and
numerical errors are under strong control, a discrepancy between results of particle and fluid model must
be attributed to the approximations made in the fluid model.

For electric fields immediately ahead of the front of 50 kV/cm or lower --- the statements hold for
nitrogen under normal condition, they are easily extended to other densities by introducing the reduced
electric field $E/N$ ---, particle and fluid model essentially agree in front speed, profile and
ionization level behind the front. When the field increases, the velocity does not vary much between
both models, but the ionization level behind the front is substantially larger, as is demonstrated for
a field of 100~kV/cm in figure~\ref{fig:1dfpcom}. In figure~\ref{fig:discrepancy} the differences between
particle and fluid model are summarized for the full field region investigated.

Can the discrepancy between the models be attributed to a particular physical mechanism and to a
particular part of the front region? In figure~\ref{fig:lfazoom} we find the largest discrepancy
between fluid and particle model in the leading edge of the front where the electron density is very
low and where the electric field is hardly screened. Here the electrons have an average energy
considerably higher than estimated by the local electric field. This effect can only be explained
by an effect of the electron density gradient, not of the electric field gradient. Indeed, an
electron swarm or avalanche in a constant electric field has the same velocity and the same density
gradients in its leading edge and shows the same electron energy overshoot in its front part,
as is illustrated in figure~\ref{fig:den_en}. In essence, in high fields the gradient length of
front and avalanche $\ell^*$ becomes of the order of the electron energy relaxation length,
therefore the fast and energetic electrons can get ahead of the slower ones.

These higher electron energies in the leading edge of the front lead to higher ionization rates
than the ionization rate $\alpha(E)$ in local field approximation --- indeed, $\alpha(E)$ is the
total effective ionization rate of a complete electron swarm (as derived in section~\ref{swarm})
while the electrons in the leading edge of the swarm have higher ionization rates.

Now the ionization level behind the front can be approximated by integrating the effective
ionization rate $\alpha(E)$ over the electric field strength $E$, independently of the precise
spatial structure of the front. This result is derived in the appendix. This implies that
higher electron energies and higher effective ionization rates in the leading edge of the front
directly lead to higher ionization levels behind the front, even though only very few electrons
are involved in this dynamics in the low density region.

The effect of few electrons of high energy is much more severe for the ionization rates $\alpha$
than for the average drift velocity $\mu E$, therefore the most pronounced effect is seen in the
ionization levels behind the front, and less in the front velocities.

We finally remark that next to explicit predictions, our work has delivered two useful insights:
$(i)$ the physical discrepancies between particle and fluid model lie in the leading edge of the front,
though the effect is not so much seen in the velocity, but much more in the ionization level
behind the front. This gives a clue for a numerical strategy combining efficient features of
fluid and particle model. $(ii)$ The essential features in the leading edge of the front are
equally present in the leading edge of an electron swarm or avalanche in a constant field, where it can be studied much easier.
\begin{acknowledgements}
The authors thank W. Hundsdorfer for helpful discussions about the numerical solution of the fluid model. They acknowledge support by the Dutch national program BSIK, in the ICT project BRICKS, theme MSV1.
\end{acknowledgements}

\begin{appendix}

\section{Approximating the ionization level behind the front}

In previous papers \cite{Ebe1997,Arr2004}, we have established that for a planar, uniformly propagating front with
field independent electron mobility $\mu(E)=$~const.\ and for vanishing electron diffusion $D(E)=0$, there is a
unique relation between the field ahead of the front $E^+$, the ionization rate function $\alpha(E)$ and the
degree of ionization $n^-_e=n_p^-$ behind the front, namely \be \label{D0} n_e^-=n_p^-=\frac{\epsilon_0}{\rm
e}\int_0^{|E^+|}\alpha(e)\;de \ee in dimensional units.

Such a simple identity does not hold for the full fluid model (\ref{eq:Poisson}), (\ref{fluid1})--(\ref{eq:np})
with electron diffusion, but we can establish the expression (\ref{D0}) as an upper bound for the free electron
density $n_e^-$ behind the front; this upper bound is actually a very good approximation to the real value as
table~\ref{tab:numeri_pf} shows. The upper bound is constructed as follows:

In the ionization source term ${\cal S}$ (\ref{S}), we follow the usual procedure to take only the drift term of
the electron current into account. We note at this place that this assumption requires some reconsideration, and
one could argue that the complete current should be taken into account and that a decomposition into a drift and a
diffusion part is artificial. This argument will have to be elaborated at some other place. We here just change
the source term into \be \label{Sj} {\cal S}_j=|{\bf j}_e|\;\alpha(E), \ee while keeping the equations unchanged
otherwise. This change of the source term means that more ionization is created at each part of the front, and
therefore the ionization level behind the front will be higher. This is because in a negative front, drift current
and diffusion current point into the same direction, and therefore \be \label{Ineq} |{\bf
j}_e|=\Big|-\underline{\underline{\bf D}}(E)\cdot\nabla n_e-\mu(E)\;{\bf E}\;n_e\Big| \gtrapprox \mu(E)\;|{\bf
E}|\;n_e. \ee The dominant part of the current is the drift or Ohmic part, therefore the results with the changed
source term (\ref{Sj}) should be close to the original problem.

According to Maxwell's equations, the divergence of the total current vanishes \be \label{J_tot}
\nabla\cdot(\epsilon_0\;\partial_t {\bf E}+{\bf j})=0, \ee where ${\bf j}$ is the electric current; in our case
it is ${\bf j} = -{\rm e}\;{\bf j}_e$ with ${\bf j}_e$ from \eqref{eq:momentum}. If the front
is planar, and if the electric field $E^+$ in the non-ionized region does not change in time, then it follows
immediately that \be \label{J} \epsilon_0\;\partial_tE={\rm e}\;j_e \ee Now the growth of the ion density
(\ref{fluid1}) is bounded by the source term (\ref{Sj}) and the identity (\ref{J_tot}) is inserted:
\begin{eqnarray}
\partial_tn_p&=&{\cal S}=\mu(E)\;|{\bf E}|\;n_e\;\alpha(E)
\nn \\
&&\lessapprox {\cal S}_j= j_e\;\alpha(E) =\frac{\epsilon_0}{\rm e}\;\partial_tE\;\alpha(E).
\end{eqnarray}
The first and last expression in this inequality can be integrated in time with the result \be
n_p({\bf x},t)-n_p({\bf x},0)\;\lessapprox\;\frac{\epsilon_0}{\rm e} \int_{E({\bf x},0)}^{E({\bf
x},t)}\alpha(e)\;de. \label{final} \ee Now the time interval $[0,t]$ is taken in such a way that it contains
the time range in which the ionization front passes over the point ${\bf x}$ of observation. $n_p({\bf x},0)$
is then the ion density before and $n_p({\bf x},t)$ the ion density behind the front. As a result, we find that the
expression (\ref{D0}) indeed is an upper bound and good approximation for the ionization level behind the front.

We finally note if there is electron attachment and positive and negative ions $n_{p,n}$ are formed, the statement
stays true for the total ion charge density $n=n_p-n_n$ if the total source and sink term for the ion density $n$
can be written in the form (\ref{Sj}).

\end{appendix}

\begin{thebibliography}{10}

\bibitem{Loe1941}
L.~B. Loeb and J.~M. Meek.
\newblock {\em The Mechanism of the Electric Spark}.
\newblock Clarendon Press, Oxford, 1941.

\bibitem{Rae1964}
H.~Raether.
\newblock {\em Electron Avalanches and Breakdown in Gases}.
\newblock Butterworths, London, 1964.

\bibitem{Lay2003}
B.~Lay, R.~S. Moss, S.~Rauf, and M.~J. Kushner.
\newblock {\em Plasma Sources Sci. Technol.}, 12:8--21, 2003.

\bibitem{Bho2004}
A.~Bhoj and M.~J. Kushner.
\newblock {\em J. Phys. D: Appl. Phys.}, 37:2510--2526, 2004.

\bibitem{Vel2000}
E.~M. van Veldhuizen, editor.
\newblock {\em Electrical Discharges for Environmental Purposes: fundamentals
  and applications}.
\newblock Nova Science Publishers, 2000.

\bibitem{Gab2005}
L.~R. Grabowski, E.~M. van Veldhuizen, A.~J.~M. Pemen, and W.~R. Rutgers.
\newblock {\em Plasma Chem. and Plasma Proc.}, 26(1):3--17, 2005.

\bibitem{Baz2000}
E.~M. Bazelyan and Yu.~P. Raizer.
\newblock {\em Lightning Physics and Lightning Protection}.
\newblock Institute of Physics, Bristol, 2000.

\bibitem{Dwy2003}
J.~R. Dwyer.
\newblock {\em Geophys. Res. Lett.}, 30(20):2055, 2003.

\bibitem{Wil2006}
E.~R. Williams.
\newblock {\em Plasma Sources Sci. Technol.}, 15(2), 2006.

\bibitem{Sen1995}
D.~D. Sentman, E.~M. Wescott, D.~L. Osborne, D.~L. Hampton, and M.~J. Heavner.
\newblock {\em Geophys. Res. Let.}, 22(10):1205--1208, 1995.

\bibitem{Dha1987}
S.~K. Dhali and P.~F. Williams.
\newblock {\em J. Appl. Phys.}, 62(12):4696--4707, 1987.

\bibitem{Guo1993}
J.~M. Guo and C.~H.~J. Wu.
\newblock {\em IEEE Trans. on Plasma Science}, 21(6):684--695, 1993.

\bibitem{Vit1994}
P.~A. Vitello, B.~M. Penetrante, and J.~N. Bardsley.
\newblock {\em Phys. Rev. E}, 49(6):5574--5589, 1994.

\bibitem{Kul1994}
A.~A. Kulikovsky.
\newblock {\em J. Phys. D: Appl. Phys.}, 27(12):2556--2563, 1994.

\bibitem{Kul1995}
A.~A. Kulikovsky.
\newblock {\em J. Phys. D: Appl. Phys.}, 28(12):2483--2493, 1995.

\bibitem{Bab1996}
N.~Yu. Babaeva and G.~V. Na\u{\i}dis.
\newblock {\em J. Phys. D: Appl. Phys.}, 29(9):2423--243, 1996.

\bibitem{Ebe1996}
U.~Ebert, W.~van Saarloos, and C.~Caroli.
\newblock {\em Phys. Rev. Lett.}, 77(20):4178--4181, 1996.

\bibitem{Ebe1997}
U.~Ebert, W.~van Saarloos, and C.~Caroli.
\newblock {\em Phys. Rev. E}, 55(2):1530--1549, 1997.

\bibitem{Mon2006:3}
C.~Montijn, W.~Hundsdorfer, and U.~Ebert.
\newblock {\em J. Computational Phys.}, 219:801--835, 2006.

\bibitem{Seg2006}
P.~S\'egur, A.~Bourdon, E.~Marode, D.~Beeieres, and J.H. Paillol.
\newblock {\em Plasma Sources Sci. Technol.}, 15:648--660, 2006.

\bibitem{Luq2007}
A.~Luque, U.~Ebert, C.~Montijn, and W.~Hundsdorfer.
\newblock {\em Appl. Phys. Lett.}, 90:in print, 2007.

\bibitem{Nai1997}
G.~V. Na\u{\i}dis.
\newblock {\em Tech. Phys. Lett.}, 23(6), 1997.

\bibitem{Kun1988:1}
E.~E. Kunhardt, J.~Wu, and B.~Penetrante.
\newblock {\em Phys. Rev. A}, 37(5):1654--1662, 1986.

\bibitem{Liu2004}
N.~Liu and V.~P. Pasko.
\newblock {\em J. Geophys. Res.}, 109:A04301, 2004.

\bibitem{Vel2002}
E.~M. van Veldhuizen and W.~R. Rutgers.
\newblock {\em J. Phys. D: Appl. Phys.}, 35:2169--2179, 2002.

\bibitem{Brie2005}
T.~M.~P. Briels, E.~M. van Veldhuizen, and U.~Ebert.
\newblock {\em IEEE Trans. on Plasma Science}, 33(2):264--265, 2005.

\bibitem{Ebe2006}
U.~Ebert, C.~Montijn, T.~M.~P. Briels, W.~Hundsdorfer, B.~Meulenbroek,
  A.~Rocco, and E.~M. van Veldhuizen.
\newblock {\em Plasma Sources Sci. Technol.}, 15:S118--S129.

\bibitem{Arr2002}
M.~Array{\'a}s, U.~Ebert, and W.~Hundsdorfer.
\newblock {\em Phys. Rev. Letters}, 88, 2002.

\bibitem{Roc2002}
A.~Rocco, U.~Ebert, and W.~Hundsdorfer.
\newblock {\em Phys. Rev. E}, 66:035102(R), 2002.

\bibitem{Mon2006:2}
C.~Montijn, U.~Ebert, and W.~Hundsdorfer.
\newblock {\em Phys. Rev. E}, 73:065401, 2006.

\bibitem{BRI2006}
T.M.P. Briels, J.~Kos, E.M. van Veldhuizen, and U.~Ebert.
\newblock {\em J. Phys. D: Appl. Phys.}, 39:5201--5210, 2006.

\bibitem{Dwy2004}
J.~R. Dwyer.
\newblock {\em Geophys. Res. Lett.}, 31:L12102, 2004.

\bibitem{Smi2005}
D.~M. Smith, L.~I. Lopez, R.~P. Lin, and C.~P. Barrington-Leigh.
\newblock {\em Science}, 307:1085--1088, 2005.

\bibitem{Dwy2005}
J.~R. Dwyer.
\newblock {\em Geophys. Res. Lett.}, 32:L20808, 2005.

\bibitem{Mos2006}
G.~D. Moss, V.~P. Pasko, Ningyu Liu, and G.~Veronis.
\newblock {\em J. Geophys. Res.}, 111:A02307, 2006.

\bibitem{Ebe2000}
U.~Ebert and W.~van Saarloos.
\newblock {\em Physica D}, 146:1--99, 2000.

\bibitem{Oli2005}
O.~Chanrion and T.~Neubert.
\newblock To be published

\bibitem{Lag1994}
A.~N. Lagarkov and I.~M. Rutkevich.
\newblock {\em Ionization Waves in Electrical Breakdown of Gases}.
\newblock Springer, Berlin, 1994.

\bibitem{Ben1978}
C.~M. Bender and S.~A. Orszag.
\newblock {\em Advanced Mathematical Methods for Scientists and Engineers}.
\newblock McGraw-Hill, New York, 1978.

\bibitem{Fif1988}
P.C. Fife.
\newblock {\em Dynamics of internal layers and diffusive interfaces}.
\newblock  SIAM, Philadelphia, 1978.

\bibitem{Ebe2000:1}
U.~Ebert and W.~van~Saarloost.
\newblock {\em Phys. Rep.}, 337:139--156, 2000.

\bibitem{Mor1999}
W.~L. Morgan, J.~P. Boeuf, and L.~C. Pitchford.
\newblock Siglo cross sections, http://www.siglo-kinema.com.

\bibitem{Kun1986:1}
E.~E. Kunhardt and Y.~Tzeng.
\newblock {\em Phys. Rev. A}, 34(3):2158--2166, 1986.

\bibitem{Okh2002}
A.~Okhrimovskyy, A.~Bogaerts, and R.~Gijbels.
\newblock {\em Phys. Rev. E}, 65:037402, 2002.

\bibitem{Phe1985}
A.~V. Phelps and L.~C. Pitchford.
\newblock {\em Phys. Rev. A}, 31(5):2932--2949, 1985.

\bibitem{Boe1982}
J.~P. Boeuf and E.~Marode.
\newblock {\em J. Phys. D: Appl. Phys.}, 15:2169--2187, 1982.

\bibitem{Sur1990}
M.~Surendra, D.~B. Graves, and G.~M. Jellum.
\newblock {\em Phys. Rev. A}, 41(2):1112--1125, 1990.

\bibitem{opa1971}
C.~B. Opal, W.~K. Peterson, and E.~C. Beaty.
\newblock {\em The Journal of Chemical Physics}, 55(8):4100--4106, 1971.

\bibitem{Bir1991}
C.~K. Birdsall and A.~B. Langdon.
\newblock {\em Plasma Physics via Computer Simulation}.
\newblock Adam Hilger, 1991.

\bibitem{Wou2005}
W.J.M. Brok.
\newblock {\em Modelling of Transient Phenomena in Gas Discharge}.
\newblock PhD thesis, Eindhoven University of Technology, The Netherlands,
  2005.

\bibitem{Hir1954}
J.~O. Hirschfelder, C.~F. Curtiss, and R.~B. Bird, editors.
\newblock {\em Molecular Theory of Gases and Liquids}.
\newblock Wiley \& Sons, New York, 1964.

\bibitem{Cha1974}
S.~Chapman and T.~G. Cowling.
\newblock {\em The Mathematical Theory of Non-Uniform Gases}.
\newblock Cambridge University Press, 1974.

\bibitem{Shk1966}
I.~P. Shkarofsky, T.~W. Johnston, and M.~P. Bachynski.
\newblock {\em The Particle Kinetics of Plasmas}.
\newblock Addison-Wesley, 1966.

\bibitem{Gog1992:1}
E.~Gogolides and H.~H. Sawin.
\newblock {\em J. Appl. Phys.}, 72(9):3971--3987, 1992.

\bibitem{sat1985}
N.~Sato and H.~Tagashira.
\newblock {\em J. Phys. D: Appl. Phys.}, 18:2451--2461, 1985.

\bibitem{Ale1996}
N.~L. Aleksandrov and I.~V. Kochetov.
\newblock {\em J. Phys. D: Appl. Phys.}, 29:1476--1483, 1996.

\bibitem{Hag2005}
G.~J.~M. Hagelaar and L.~C. Pitchford.
\newblock {\em Plasma Sources Sci. Technol.}, 14(4):722--733, 2005.

\bibitem{Mon2006:1}
C.~Montijn and U.~Ebert.
\newblock {\em J. Phys. D: Appl. Phys.}, 39:2979--2992, 2006.

\bibitem{Hux1974}
L.~G.~H. Huxley and R.~W. Crompton.
\newblock {\em The diffusion and drift of electrons in gases}.
\newblock Wiley, New York, 1974.

\bibitem{Rai1991}
Yu.~P. Raizer.
\newblock {\em Gas Discharge Physics}.
\newblock Springer, Berlin, 1991.

\bibitem{Arr2004}
M.~Array{\'a}s and U.~Ebert.
\newblock {\em Phys. Rev. E}, 69:036214, 2004.

\end{thebibliography}

\end{document}